\documentclass[11pt]{article}
\usepackage{amssymb,amsmath}
\usepackage{bm} 
\usepackage{booktabs} 
\usepackage{array}
\usepackage{latexsym}
\usepackage{graphicx}
\usepackage{color}
\usepackage{datetime}
\usepackage[nosort]{cite}
\usepackage{verbatim}
\usepackage{enumerate}
\usepackage{chngpage} 
\usepackage{mathrsfs}
\usepackage{euscript}
\usepackage{psfrag}

\usepackage{mciteplus}

\usepackage[colorlinks=true,      linkcolor=blue,      urlcolor=blue,      
            filecolor=blue,      citecolor=blue,       pdfstartview=FitH,     
						pdfpagemode=UseNone,      bookmarksopen=true]{hyperref}  
\usepackage[all]{hypcap}     

\def\eq#1{(\ref{#1})}

\newcommand{\comm}[1]{} 

\def\({\left(}
\def\){\right)}
\def\[{\left[}
\def\]{\right]}

\def\coeff#1#2{{\textstyle \frac{#1}{#2}}}

\def\One{{\hbox{ 1\kern-.8mm l}}}

\def\barray{\begin{array}}
\def\earray{\end{array}}
\def\be{\begin{equation}}
\def\ee{\end{equation}}
\def\bea{\begin{eqnarray}}
\def\eea{\end{eqnarray}}
\def\bal{\begin{align}}
\def\eal{\end{align}}

\numberwithin{equation}{section} 

\makeatletter
\g@addto@macro\bfseries{\boldmath}
\makeatother

\definecolor{cardinal}{rgb}{0.6,0,0}
\definecolor{darkgreen}{rgb}{0,0.4,0}
\definecolor{golden}{rgb}{0.92, 0.7, 0}
\definecolor{midnight}{rgb}{0, 0, 0.5}
\definecolor{darkblue}{rgb}{0, 0, 0.7}
\definecolor{purple}{rgb}{0.5, 0, 0.5}

\def\Neql#1{{\cal N}\!=\!{#1}}

\def\IR{\mathbb{R}}

\def\IT{\mathbb{T}}

\def\cB{{\cal B}}

\def\cD{{\cal D}}
\def\cF{{\cal F}}

\def\cL{{\cal L}}

\def\cN{{\cal N}}
\def\cP{{\cal P}}

\def\cW{{\cal W}}

\def\nBPS#1{$\frac{1}{#1}$-BPS}

\begin{document}

\phantom{AAA}
\vspace{-10mm}

\begin{flushright}
IPHT-T19/021\\
\end{flushright}

\vspace{1.9cm}

\begin{center}

{\huge {\bf Superstratum Symbiosis}}

\vspace{1cm}

{\large
\textsc{Pierre Heidmann$^1$} and \textsc{Nicholas P.~Warner$^{1,8}$}}

\vspace{1cm}

$^1$Institut de Physique Th\'eorique,
Universit\'e Paris Saclay,\\
CEA, CNRS, F-91191 Gif sur Yvette, France \\
\medskip
$^8$Department of Physics and Astronomy
and Department of Mathematics,\\
University of Southern California,
Los Angeles, CA 90089, USA

\vspace{10mm} 
{\footnotesize\upshape\ttfamily pierre.heidmann @ ipht.fr, warner @ usc.edu} \\

\vspace{2.2cm}
 
\textsc{Abstract}

\end{center}

\begin{adjustwidth}{3mm}{3mm} 
 
\vspace{-1.2mm}
\noindent
 Superstrata are smooth horizonless microstate geometries for the supersymmetric D1-D5-P black hole in type IIB supergravity. In the CFT,  ``superstratum states'' are defined to be the component of the supergraviton gas that is obtained by breaking the CFT into ``$|00\rangle$-strands'' and acting on each strand with  the ``small,'' anomaly-free superconformal generators. We show that the recently-constructed supercharged superstrata represent a final and crucial component for the construction of the supergravity dual of a generic superstratum state and how the supergravity solution faithfully represents all the coherent superstratum states of the CFT.  For the supergravity alone, this shows that generic superstrata do indeed fluctuate as functions of {\it three independent variables}.   Smoothness of the complete supergravity solution also involves ``coiffuring constraints'' at second-order in the fluctuations and we describe how these lead to new predictions for three-point functions in the dual CFT. We use a hybrid of the original and supercharged superstrata to construct families of single-mode superstrata that still have free moduli after one has fixed the asymptotic charges of the system. We also study scalar wave perturbations in a particular family of such solutions and show that the mass gap depends on the free moduli. This can have interesting implications for superstrata at non-zero temperature.

\end{adjustwidth}

\thispagestyle{empty}
\newpage


\baselineskip=17pt
\parskip=5pt

\setcounter{tocdepth}{2}
\tableofcontents

\baselineskip=15pt
\parskip=3pt

\section{Introduction}
\label{sec:Intro}

The construction of superstrata  \cite{Bena:2015bea} has proven to be a very important advance in the development of microstate geometries.  In particular, it has led to the first, smooth, horizonless  geometries in six (and five) dimensions that closely approximate macroscopic three-charge black holes with arbitrarily small angular momentum \cite{Bena:2016ypk,Bena:2017geu}.  These microstate geometries can be arranged to be asymptotically flat \cite{Bena:2017xbt}, or asymptotic to AdS$_3 \times S^3$  \cite{Bena:2015bea}, and have arbitrarily long BTZ-like throats, in which the background geometry behaves like that of AdS$_2 \times S^1 \times S^3$, and yet this throat ultimately caps off in smooth geometry at high red-shift.  The holographic dictionary for these geometries is also sufficiently well-developed to identify the holographically dual states in the D1-D5 CFT and determine precisely what part of the black-hole microstate structure is being captured by the superstratum.

The underlying system starts with $N_5$ D5 branes wrapped on a ${\Bbb T}^4 \times S^1$ (or K3 $\times S^1$) and $N_1$ D1 branes wrapped on the $S^1$.  The CFT world-volume is the common $S^1$ of the D1's and D5's.   Adding angular momentum and back-reacting the branes  leads to the simplest of microstate geometry:  the \nBPS{4}, D1-D5 supertube of IIB supergravity.  The superstratum may be then viewed as a generalization of the supertube that involves adding not just shape modes to the supertube but adding left-moving momentum charge along the common  $S^1$ of the D1-D5 system.    This can be arranged so that it back-reacts to produce a smooth, three-charge (D1-D5-P),  \nBPS{8} microstate geometry that is dual to some of the microstate structure of the black hole.

An equivalent way to think of the superstratum is as a fluctuating, non-trivial topological $3$-cycle in six dimensions.  The  $3$-cycle can be described as the common $S^1$ of the D1-D5 system fibered over the disk spanning a simple, closed supertube loop.  The common $S^1$ actually pinches off at the supertube locus and thus creates a $3$-cycle.  This cycle is held up my the fluxes sourced by the D1 and D5 branes and, without fluctuations, this is precisely the smooth resolution of the supertube geometry into an AdS$_3 \times S^3$ geometry. This geometry is simply the holographic dual of a D1-D5 CFT ground state.  The superstratum involves adding fluctuations to this background  and computing the back-reacted geometry that these fluctuations produce.  The fluctuations are, of course, dual to excited states of the CFT and, if they are purely left-moving, one gets the \nBPS{8} superstratum. 

To construct ``superstratum states'' in the CFT one uses the generators of the ``small,'' anomaly-free $\Neql{4}$ superconformal algebra  to act on a particular ``length-$k$'' strand in the  NS-NS ground state of the D1-D5 CFT :
\begin{equation}
\begin{aligned}
 |k,m,n,q\rangle^{\rm NS} &=
 \begin{cases}
 (J^+_0)^{m} (L_{-1})^{n}  \,  |O^{--}\rangle_k \,, \qquad q=0  \\  
  (J^+_0)^{m-1} (L_{-1})^{n-1}  \left(G_{-\frac12}^{+1}G_{-\frac12}^{+2} + \frac{1}{k} J^+_0 L_{-1}\right)  |O^{--}\rangle_k \,, \qquad q=1  \,,
\end{cases} \\
& ~=~ (J^+_0)^{m-1} (L_{-1})^{n-1}  \bigg(q \, G_{-\frac12}^{+1}G_{-\frac12}^{+2} + \Big(1 + \frac{q(1-k)}{k}\Big)\, J^+_0 L_{-1}\bigg)\, |O^{--}\rangle_{k} \,,
\end{aligned}
\label{SSCFTstates}
\end{equation} 
where $q=0,1$; $n \ge 1$ and $k >0, k -q \ge m \ge 1$.   For more details, see \cite{Ceplak:2018pws}. One should note that we are labelling the states slightly differently than in \cite{Ceplak:2018pws}.  We do this because our labelling conventions, (\ref{SSCFTstates}), line up better with eigenvalues of operators and with the supergravity mode numbers. A basis for the superstratum states is then obtained by taking  tensor products of such excitations\footnote{This form of the superstratum state is a natural extrapolation of the states for $q=0$ and $q=1$, however there may be some subtlety in determining whether the coherent state is peaked over this state, or a superposition of states like this.  We leave this to future work as it does not influence the superstrata that we construct here.}:
\begin{equation}
\begin{aligned}
 \bigotimes_{i}\, & |k_i,m_i,n_i,q_i\rangle^{\rm NS}  \\
 &~=~  \bigotimes_{i}\,  \bigg[(J^+_0)^{m_i-1} (L_{-1})^{n_i-1}  \bigg(q_i \, G_{-\frac12}^{+1}G_{-\frac12}^{+2} + \Big(1 + \frac{q_i(1-k)}{k}\Big)\, J^+_0 L_{-1}\bigg)\, |O^{--}\rangle_{k_i}\bigg]\,,
 \end{aligned}
   \label{multistates}
\end{equation} 
where repetitions of quantum numbers are allowed.  Indeed, we will denote the number of states with quantum numbers $(k,m,n,q)$ by $N_{k,m,n,q}$.   Since the total strand-length is limited by the size of the chiral ring (or degeneracy of the R-R ground states), there is a limitation on the ``strand-budget:''
\begin{equation}
\sum_{i}  \, k_i \, N_{k_i,m_i,n_i,q_i} ~\le~  N_1 N_5\,,
   \label{sbudget}
\end{equation} 

In supergravity, the duals of these excited CFT states appear, at leading order, in the fluctuations of a particular tensor gauge field usually denoted by $\Theta_4$.   The quantum numbers, $(k_i,m_i,n_i)$, in the CFT state correspond, in supergravity, to  mode numbers of the fluctuations while $q_i = 0,1$, determines the particular components of $\Theta_4$ that are being excited.   The  ``occupation numbers,'' $N_{k_i,m_i,n_i,q_i}$, are related to the Fourier coefficients, $b_{k_i,m_i,n_i,q_i}$, of these leading modes via:
\begin{equation}
N_{k_i,m_i,n_i,q_i} ~\sim~  (b_{k_i,m_i,n_i,q_i})^2\,.
   \label{NandFCnorms}
\end{equation} 
Since the occupation numbers are free parameters in the CFT (modulo the bound  (\ref{sbudget})), a complete holographic correspondence requires that  the Fourier coefficients, $b_{k_i,m_i,n_i,q_i}$, must also be free parameters of the superstratum (modulo  the supergravity analog of  (\ref{sbudget})). 

In the fully back-reacted supergravity solution, these ``leading modes'' interact with one another to produce new ``secondary'' excitations in the metric and in the other tensor gauge fields.  In finding the supergravity solutions, such secondary excitations are initially incorporated using arbitrary Fourier coefficients.  However, smoothness of the final solution leads to ``coiffuring constraints'' \cite{Bena:2013ora,Bena:2014rea,Bena:2015bea,Bena:2017xbt} that generically determine all the secondary Fourier coefficients in terms of the leading Fourier coefficients.    

Such coiffuring constraints also have direct counterparts in the dual CFT.  The secondary excitations are dual to light operators and their Fourier coefficients give the vevs of these operators in the ``heavy state'' corresponding to the full supergravity solution (see, for example, \cite{Giusto:2015dfa}).  A coiffuring constraint thus leads to a definite prediction for the vev of a particular operator in the dual CFT state.  Such ``precision holography'' is an immensely powerful tool in testing the holographic dictionary.  For superstrata, many of the coiffuring constraints are automatically satisfied through the solution generating methods that were used to obtain the supergravity excitations.  However, there are some Fourier coefficients that are not fixed in this way but are determined, in supergravity, only through smoothness. Such constraints can thus be used to perform non-trivial tests with precision holography \cite{Kanitscheider:2006zf,Kanitscheider:2007wq,Giusto:2015dfa} and in \cite{Giusto:2015dfa} it was shown that a particular class of superstratum coiffuring constraints matched perfectly between supergravity and CFT. 

This illustrates a very broad thread in supergravity and holographic CFT's:  The correct description of the infra-red physics of the CFT, and most particularly the vevs of order parameters of the infra-red phase, become manifest in supergravity through the smoothness  of the geometry.  We have seen a multitude of examples of this, ranging from the holography of confining phases of QCD \cite{Klebanov:2000hb}, through fermion droplet models \cite{Lin:2004nb} to the basic underlying philosophy of microstate geometries\cite{Bena:2007kg,Bena:2013dka}.   It is very encouraging to confirm that superstrata are not only modeling black-hole structure at a macroscopic level but also at the level of detail accessible with precision  holography.

One of the purposes of this paper is to use smoothness constraints in superstrata to make a new class of predictions about three-point functions, or vevs, that can be thoroughly tested within the CFT.  Indeed, as we will describe below, we arrive at these predictions by using the new supercharged superstrata to solve an outstanding problem in the original construction of families of  superstrata. 

The ``original superstrata,'' which have been extensively studied over the last four years \cite{Bena:2015bea,Bena:2016agb,Bena:2016ypk,Bena:2017geu,Bena:2017upb,Tyukov:2017uig,Bena:2017xbt,Bena:2018bbd,Tyukov:2018ypq,GreenSuperstrata}, were restricted to the action of the bosonic sub-algebra, that is, to the states with $q_i=0$.   Moreover, much of the focus  has been on classes of ``single-mode'' superstrata that are dual to excitations of a single state of the form (\ref{SSCFTstates}). Such geometries have the virtue of being relatively simple and yet exhibit an incredibly rich structure in terms of parallels with, and differences from, black holes.  They have also revealed a lot of interesting physics encoded in microstate geometries.  

In this paper, we use the new ``supercharged'' single-mode superstrata, with $q=1$, to construct the \emph{most general  ``single-mode" superstratum}. That is,  we will make a ``hybrid" of the original single-mode superstratum, dual to  $|k,m,n,q =0 \rangle^{\rm NS}$, and the supercharged superstratum state, dual to $|k,m,n,q=1 \rangle^{\rm NS}$: $$|k,m,n,q =0 \rangle^{\rm NS} \otimes |k,m,n,q=1 \rangle^{\rm NS}. $$  This will still be a ``single-mode'' superstratum in that it only depends on one Fourier mode.  However the two distinct states enter the supergravity through different tensor structures in $\Theta_4$ and thus are represented by two distinct Fourier coefficients.  These two tensor structures are extraordinarily miscible and their combined back-reaction remains relatively simple.  In particular, the fluctuating modes only contribute to the background metric via their RMS values.  

One of the down-sides of the original ``single-mode'' superstrata was that there were no free moduli. Such solutions have a significant number of free parameters, but once smoothness is imposed, the remaining parameters are completely determined by the bulk thermodynamic variables:  the charges and the angular momenta of the system.   This means that one is working with a single, highly coherent pure state of the CFT and such objects can have some very ``atypical'' properties and there is no way to study how such atypicality might ``wash-out'' in  an average over more generic states of the form (\ref{multistates}).  Our new ``hybrid'' single-mode superstrata have a modulus that varies the amount of each species of state.  It thus affords the possibility of studying how the physical properties  of superstrata vary across a family of states with the same values of the bulk thermodynamic variables. 

As we described in detail above, there must be a lot more moduli in the  general multi-mode superstrata that are dual to states of the form (\ref{multistates}).  There are two issues with this.  The most obvious practical problem is the complexity of the solution.   However, it was shown in \cite{Bena:2015bea} that there are some rich families of superstrata that are dual to some families of states of the form (\ref{multistates}) (with $q_i=0$).  These solutions are complicated but one can build up some classes of multi-mode superstrata algorithmically.

The second issue is much more fundamental and more technical: resolving all the constraints imposed by the smoothness of the supergravity solutions.    As we indicated above, such coiffuring constraints fix the secondary Fourier coefficients in terms of the  ``leading''  coefficients.  However, it became apparent that for the original classes of superstrata ($q_i=0$), regularity of completely general multi-mode excitations would place some further restrictions on the ``leading'' Fourier coefficients as well.  This would suggest a major deficiency in  superstrata as duals of the CFT states. In supergravity, these additional constraints also suggested that generic superstrata might only depend on certain two-dimensional subspaces of modes, rather than on  fully general functions of three variables.  However, we will show that this constraint problem is completely resolved with the hybrid superstrata and that a generic superstratum fluctuations are indeed described by two unconstrained functions of three variables.

More specifically, we will show that including the supercharged  ($q=1$)  superstratum states leads us to introduce  new ``leading'' and ``secondary'' Fourier modes into the supergravity solution.  In particular, the additional secondary Fourier coefficients allow us to solve the essential coiffuring constraints and leave us with precisely the correct set of free leading  Fourier coefficients, $b_{k_i,m_i,n_i,q_i}$, that are the duals of the freely choosable  $N_{k_i,m_i,n_i,q_i}$  in the CFT.  We thus have exactly the right number of free parameters to describe a generic multi-moded superstratum that is a function of three independent variables and is dual to a generic state of the form  (\ref{multistates}).  There is thus no deficiency in superstrata when it comes to describing the coherent duals of states of the form (\ref{multistates}).   This resolution is achieved by a constraint that involves both ``leading'' modes of the original superstrata and ``secondary'' modes of  supercharged superstrata.  Thus the supergravity leads to new holographically testable CFT predictions:   interactions of the original superstratum states lead to  excitations of light supercharged operators, and their expectation values are determined by the coiffuring constraints.

Finally, in the interest of full disclosure, we should point out that the claim about  coiffuring and the residual free coefficients, $b_{k_i,m_i,n_i,q_i}$, is not at the level of a rigorous mathematical theorem, but is based upon  experience in computing many explicit examples. The sources of the BPS equations are quadratic in the mode expansions and so a generic multi-mode solution simply involves many two-mode interactions.  In each such interaction, the mode numbers either ``add'' to create ``high-frequency'' sources, or subtract to create ``low-frequency'' sources.  Experience shows that it is only the  high-frequency sources that lead to singular solutions, and so it is only these sources that need coiffuring\footnote{The designation of high and low frequency modes is largely unambiguous because the mode numbers are non-negative.  When a non-trivial mode interacts with a zero mode, this is viewed as a high-frequency source. When two zero modes interact, the result is ``low-frequency'' in that it can easily be handled and arranged to produce non-singular solutions.}.  In this paper we will show that one can indeed coiffure these ``high-frequency'' sources and still leave precisely the right number of free parameters to describe all the CFT states in  (\ref{multistates}).
  
  In Sections \ref{Sect:sixD},  \ref{sec:OriginalSuperstrata} and  \ref{sec:SuperchargedSuperstrata}   we review the essential elements of six-dimensional supergravity, the original superstrata and the supercharged superstrata.  In Section  \ref{sec:Hybrids} we construct the most general single-mode hybrid superstratum and show that it still has moduli after the asymptotic charges are fixed.   In Section \ref{sec:example} we give even more details for the $(2,1,n)$ single-mode, investigating the geometry near the cap of the superstratum and modes scalar Laplacian in various limits. Section \ref{sec:multiSS} contains our discussion of the multi-mode superstrata and how the supercharged modes enable coiffuring of generic multi-mode superstrata and lead to predictions of three-point functions in the CFT.  We then make some final remarks in Section \ref{sec:Conclusions}.


\section{Superstrata in supergravity}
\label{Sect:sixD}

Superstrata are  most simply described within the six-dimensional $(0,1)$ supergravity obtained by compactifying  IIB supergravity  on ${\Bbb T}^4$ (or $K3$) and then truncating the matter spectrum to tensor multiplets.   Indeed, one typically works within the simplest class of models that can lead to (smooth) superstrata:  $(0,1)$ supergravity coupled to two tensor multiplets.   For the BPS solutions, the six-dimensional metric takes the form  \cite{Gutowski:2003rg,Cariglia:2004kk}:
\begin{equation}
d s^2_{6} ~=~-\frac{2}{\sqrt{\cP}}\,(d v+\beta)\,\Big[d u+\omega + \frac{\mathcal{F}}{2}(d v+\beta)\Big]+\sqrt{\cP}\,d s^2_4\,,
\label{sixmet}
\end{equation} 
where
\begin{equation}
u ~\equiv~ \frac{1}{\sqrt{2}}(t-y)\,, \qquad \qquad v~\equiv~\frac{1}{\sqrt{2}}(t+y)
\label{uvdefn}
\end{equation} 
are null coordinates and $y$ parametrizes the common $S^1$ of the D1 and D5 branes.  

For the standard classes of superstrata, the metric, $ds^2_4$, on the four-dimensional spatial base, $\cB$, is simply that of flat $\IR^4$ and it is most convenient to use spherical bipolar coordinates in which the  metric becomes:
\begin{equation}
 d s^2_4 = (r^2+a^2 \cos^2\theta)\, \left(\frac{d r^2}{r^2+a^2}+ d\theta^2\right)+(r^2+a^2)\sin^2\theta\,d\phi^2+r^2 \cos^2\theta\,d\psi^2\,.
\label{ds4flat}
\end{equation} 
The vector, $\beta$, is chosen to be the potential for a self-dual magnetic field on the  $\IR^4$ with a source along the supertube ring, $r=0, \theta = \frac{\pi}{2}$: 
\begin{equation}
\beta ~=~  \frac{R\,a^2}{\sqrt{2}\,\Sigma}\,(\sin^2\theta\, d\phi - \cos^2\theta\,d\psi) \,, \qquad  \Theta_3   ~\equiv~ d\beta \,, \qquad  *_4 \Theta_3 ~=~  \Theta_3 \,, 
\label{betadefn} 
\end{equation}
where $*_4$ denotes the four-dimensional dual in the metric, $ds^2_4$ and 
\begin{equation}
\Sigma ~\equiv~ (r^2+a^2 \cos^2\theta) \,. 
\label{Sigdefn} 
\end{equation}

The remaining pieces of  (\ref{sixmet}), namely the vector  $\omega$ (whose components lie on $\cB$), the functions $\cP$ and $\cF$, and the three independent tensor gauge fields, are all determined a system of BPS equations.  For  BPS solutions, each  tensor gauge field can be decomposed into an electric potential function, $Z_I$, and a magnetic $2$-form field, $\Theta_I$, on the four-dimensional base, $\cB$.  (For historical reasons, the index $I$ takes the values $1,2$ and $4$.) The details of this decomposition may be found in \cite{Bena:2011dd}, with appropriate generalization in Appendix A of  \cite{Bena:2017geu}.  

The fields and the metric in the six-dimensional supergravity encode a complete  solution to IIB supergravity, replete with all its tensor gauge field.  The ``uplift formulae'' that give the ten-dimensional solution may be found in \cite{Giusto:2013rxa, Giusto:2013bda,Bena:2015bea} and we will not repeat them here.

\noindent To write the BPS equations, we introduce the covariant exterior derivative:
\begin{equation}
\cD \equiv d- \beta \wedge \frac{\partial}{\partial v}, 
\end{equation}
where $d$ denotes the exterior derivative on the base space, $\cB$. 
The first layer of the BPS system is then
\begin{subequations}\label{layer1}
\begin{align}
*_4\cD\dot{Z}_1 = \cD \Theta_2, \qquad \cD *_4 \cD Z_1 = - \Theta_2 \wedge d\beta,\qquad \Theta_2 = *_4 \Theta_2,\\
*_4\cD\dot{Z}_2 = \cD \Theta_1, \qquad \cD *_4 \cD Z_2 = - \Theta_1 \wedge d\beta,\qquad \Theta_1 = *_4 \Theta_1,\\
*_4\cD\dot{Z}_4 = \cD \Theta_4, \qquad \cD *_4 \cD Z_4 = - \Theta_4 \wedge d\beta,\qquad \Theta_4 = *_4 \Theta_4,
\end{align}
\end{subequations}
where $\dot{}$ denotes $\frac{\partial}{\partial v}$.
This system is linear in the pairs the unknown fields $(Z_1,\Theta_2$), $(Z_2,\Theta_1$) and $(Z_4,\Theta_4$).    These functions and $2$-forms determine the tensor gauge fields and the warp factor, $\cP$: 
\begin{equation}
\cP~=~ Z_1\,Z_2 ~-~ Z_4^2.
\label{Pdefn}
\end{equation}
The momentum-potential, $\cF$, and  the angular-momentum vector, $\omega$, are determined by the second  layer of the BPS system:
\begin{align}
&\cD \omega + *_4 \cD \omega +\cF d\beta = Z_1 \Theta_1 + Z_2 \Theta_2 - 2 Z_4 \Theta_4,\nonumber\\
&*_4 \cD *_4\left(\dot{\omega}- \frac12 \cD \cF\right) = \partial_v^2(Z_1 Z_2 - Z_4^2)- (\dot{Z}_1\dot{Z_2}- (\dot{Z}_4)^2)- \frac12 *_4(\Theta_1\wedge \Theta_2 - \Theta_4 \wedge \Theta_4).\label{layer2eq}
\end{align}
In particular, note that these equations are again linear but the sources are quadratic in the solutions of the first layer of BPS equations.

\section{Original single-mode superstrata}
\label{sec:OriginalSuperstrata}

We briefly review the original single-mode superstrata, which correspond to the ($k,m,n,q=0$) superstrata states. We refer the  reader to  \cite{Giusto:2013bda, Bena:2015bea,Bena:2017xbt} for an exhaustive discussion of these solutions and their  CFT duals. Here we simply focus on the supergravity aspects of these solutions.

\subsection{The D1-D5 supertube}
\label{sec:stube}

The starting point is the 2-charge, round supertube. This has 
\begin{equation}
 \label{eq:stube1}
Z_1 ~=~  \frac{Q_1}{\Sigma} \,, \qquad Z_2  ~=~  \frac{Q_5}{\Sigma} \,,  \qquad Z_4  ~=~ 0 \,,   \qquad  \Theta_I ~=~ 0 \,. 
\end{equation}
with the non-trivial angular momentum vector, and no momentum charge
\begin{equation}
 \label{eq:stube2}
\omega~=~ \omega_0 ~\equiv~  \frac{R\,a^2}{\sqrt{2}\,\Sigma}\,(\sin^2\theta\, d\phi + \cos^2\theta\,d\psi) \,, \qquad \cF ~=~ 0 \,. 
\end{equation}
These quantities are harmonic functions and forms on $\IR^4$ and trivially satisfy the BPS equations.
The angular momentum ``spins'' the configuration up into a supertube and the metric is regular if the supertube constraint is satisfied: 
\begin{equation}
\frac{Q_1 Q_5}{R_y^2} ~=~  a^2.  
\end{equation}
The resulting smooth geometry is simply AdS$_3\times$S$^3$ with radius $(Q_1 Q_5)^{\frac{1}{4}}$.  The angular momenta are given by: 
\begin{equation}
J_L ~=~  J_R~=~    \frac{1}{2} \, \cN  \,a^2 ~=~    \frac{1}{2} \, N_1 \, N_5  \,a^2 \,, 
\label{angmom}
\end{equation}
where
\begin{equation}
\cN ~\equiv~ \frac{N_1 \, N_5\, R_y^2}{Q_1 \, Q_5} ~=~\frac{V_4\, R_y^2}{ (2\pi)^4 \,g_s^2 \,\alpha'^4}~=~\frac{V_4\, R_y^2}{(2\pi)^4 \, \ell_{10}^8} ~=~\frac{{\rm Vol} (T^4) \, R_y^2}{ \ell_{10}^8} \,.
\label{cNdefn}
\end{equation}
As  (\ref{angmom}) implies, this is dual the maximally spinning vacuum state consisting of $N_1 N_5$ strands of length $1$ in the state $|++\rangle_1$.

\subsection{The first layer}
\label{sec:oldfirst}

To make the original single-mode superstrata one first deforms the supertube by trading sets of $k$  $|++\rangle_1$ strands for  $|00\rangle_k$ strands of length $k$.  These $|00\rangle_k$ strands are then excited by acting  $m$-times, $m\leq k$, and $n$-times with the bosonic symmetry generators $J_{0}^+$ and $L_{-1}$, respectively, as  indicated in the $(k,m,n,q=0)$ states of (\ref{SSCFTstates}).   The single-mode superstratum involves a single set of modes $(k,m,n)$, which means a uniform action of $(J_{0}^+)^m (L_{-1})^n$ on (possibly multiple sets of)  $|00\rangle_k$ strands, all of the same  length $k$. In the supergravity dual, the non-trivial excitations in the first layer of the BPS system may be written in terms of the scalars and tensors:
\begin{equation}
\begin{aligned}
 \widetilde{z}_{k,m,n} &=\,R_y \,\frac{\Delta_{k,m,n}}{\Sigma}\, \cos{v_{k,m,n}},\nonumber \\
 \widetilde{\vartheta}_{k,m,n}&=-\sqrt{2}\,
\Delta_{k,m,n}
\biggl[\left((m+n)\,r\sin\theta +n\left({m\over k}-1\right){\Sigma\over r \sin\theta}  \right)\Omega^{(1)}\sin{v_{k,m,n}} \\
&\hspace{20ex}
 +\left(m\left({n\over k}+1\right)\Omega^{(2)} +\left({m\over k}-1\right)n\, \Omega^{(3)}\right) \cos{v_{k,m,n}} \biggr]\nonumber
\,,
\end{aligned} 
\label{zthetatilde}
\end{equation}
where
\begin{equation}
\begin{aligned}
  \Delta_{k,m,n} &~\equiv~
 \left(\frac{a}{\sqrt{r^2+a^2}}\right)^k
 \left(\frac{r}{\sqrt{r^2+a^2}}\right)^n 
 \cos^{m}\theta \, \sin^{k-m}\theta \,, 
 \\
 v_{k,m,n} &~\equiv~ (m+n) \frac{\sqrt{2}\,v}{R_y} + (k-m)\phi - m\psi \,,
\end{aligned} 
\label{Delta_v_kmn_def}
\end{equation}
and the $\Omega$'s are a basis of the self-dual $2$-forms: 
\begin{equation}
\label{selfdualbasis}
\begin{aligned}
\Omega^{(1)} &~\equiv~ \frac{dr\wedge d\theta}{(r^2+a^2)\cos\theta} + \frac{r\sin\theta}{\Sigma} d\phi\wedge d\psi\,,\\
\Omega^{(2)} &~\equiv~  \frac{r}{r^2+a^2} dr\wedge d\psi + \tan\theta\, d\theta\wedge d\phi\,,\\
 \Omega^{(3)} &~\equiv~ \frac{dr\wedge d\phi}{r} - \cot\theta\, d\theta\wedge d\psi\,.
\end{aligned}
\end{equation}
We will adopt the notational convention that the  quantities associated with original superstrata  will be denoted with a `` $\widetilde{\phantom{a}}${ }'' in order to distinguish them from the quantities involved in the supercharged superstrata and our new  hybrid  superstrata.
It is straightforward to check that the fields \eqref{zthetatilde} satisfy the first layer of BPS equations \eqref{layer1} 
\begin{equation}
*_4\cD\,\dot{\widetilde{z}}_{k,m,n} = \cD\,\widetilde{\vartheta}_{k,m,n}, \qquad \cD *_4 \cD\, \widetilde{z}_{k,m,n} = - \widetilde{\vartheta}_{k,m,n} \wedge d\beta,\qquad \widetilde{\vartheta}_{k,m,n} = *_4 \widetilde{\vartheta}_{k,m,n},\\
\end{equation}
The first-layer fields may then be written:
\begin{align}
 \label{eq:ZThetasinglemodeOrig}
 \begin{aligned}
 \widetilde{Z}_1 &~=~ \frac{Q_1}{\Sigma} + b_1 \,\frac{R_y }{2\, Q_5} \, \widetilde{z}_{2 k,2m,2n}\,, \qquad    \widetilde{\Theta}_2  ~=~  b_1\, \frac{R_y}{2\, Q_5}\, \widetilde{\vartheta}_{2k,2m,2n} \,; \\
 \widetilde{Z}_2 &~=~\frac{Q_5}{\Sigma} \,,  \qquad \widetilde{\Theta}_ 1  ~=~ 0\,; \qquad\qquad \widetilde{Z}_4  ~=~  b_4 \, \widetilde{z}_{k,m,n} \,, \qquad \widetilde{\Theta}_4 ~=~   b_4\, \widetilde{\vartheta}_{k,m,n} \,,
 \end{aligned}
\end{align}
where $b_1$ and $b_4$ are, as yet, arbitrary Fourier coefficients.  Note that we have only excited the pairs $(Z_1,\Theta_2$) and $(Z_4,\Theta_4$), and that 
the former has twice the mode numbers of the latter.  As we will see, this choice simplifies the second layer and makes the coiffuring is relatively easy.  The modes in $(Z_1,\Theta_2)$ represent the secondary modes described in the introduction and will be fixed  by the smoothness of the supergravity solution.  

\subsection{The second layer}
\label{sec:2ndLayOrig}

To solve the second layer we simply follow the discussion of \cite{Bena:2017xbt}. 
The sources have both a non-oscillating ``RMS'' component and an oscillating part that depends only upon $v_{2k,2m,2n}$.  Experience shows that such oscillating sources generically lead to singular angular momentum vectors and so we remove these terms by  ``coiffuring.'' That is, the Fourier coefficient of the oscillating source is proportional to $b_1 - b_4^2$ and so we take:
\begin{equation} 
b_1 ~=~b_4^2\,.
\label{origcoiff}
\end{equation} 
The solution for $\omega$ and $\cF$ is now given by the sums of the seed supertube solutions and the solution for the new pieces:
\begin{equation} 
\omega ~=~ \omega_0  ~+~ \widetilde{\omega}_{k,m,n}\,,  \qquad \cF ~=~0 ~+~\widetilde{\mathcal{F}}_{k,m,n} \,.
\label{omegaorig}
\end{equation} 
The equations \eqref{layer2eq} for $\widetilde{\omega}_{k,m,n}$ and $\widetilde{\mathcal{F}}_{k,m,n}$ reduce to
\begin{align}
&d \widetilde{\omega}_{k,m,n} + *_4 d \widetilde{\omega}_{k,m,n} +\widetilde{\mathcal{F}}_{k,m,n}\,d\beta
=
\sqrt{2} \,b_{4}^2\,R_y\, \frac{\Delta_{2k,2m,2n}}{\Sigma} \left(
\frac{m(k+n)}{k} \,\Omega^{(2)} - \frac{n(k-m)}{k} \,\Omega^{(3)} \right) , \nonumber\\
& \mathcal{L} \,  \widetilde{\mathcal{F}}_{k,m,n}
=
\frac{4\,b_{4}^2}{(r^2+a^2)\cos^2\theta \, \Sigma} 
\left[ \left( \frac{m(k+n)}{k} \right)^{2} \Delta_{2k,2m,2n} 
+ \left( \frac{n(k-m)}{k} \right)^{2} \Delta_{2k,2m+2,2n-2} \right] ,
\label{eq:F&omegaEqorig}
\end{align}
where $\mathcal{L}$ is the scalar Laplacian on the base space $\cB$, $\cL = - *_4 d *_4d$:
\begin{equation}
\mathcal{L} F \equiv \frac{1}{r\Sigma}\, \partial_r \big( r (r^2 + a^2) \, \partial_r F  \big)  +    \frac{1}{\Sigma\sin \theta \cos \theta}\partial_\theta \big( \sin \theta \cos \theta\, \partial_\theta F  \big)\,.
\label{eq:Laplace3}
\end{equation}
Note that these $\widetilde{\mathcal{F}}_{k,m,n}$ and $\widetilde{\omega}_{k,m,n}$ depend on the mode numbers but do not fluctuate themselves: coiffuring means that the metric only responds to the RMS values of the fluctuating modes.  

Since the right-hand side of the first line in \eq{eq:F&omegaEqorig} has no component in the $\Omega^{(1)}$ direction, we can set the legs of $\widetilde{\omega}_{k,m,n}$ along $dr$ and $d\theta$ to be zero and take:
\begin{equation}
\widetilde{\omega}_{k,m,n} \;\equiv\;  \widetilde{\mu}_{k,m,n} \,(d\psi+d\phi) + \widetilde{\zeta}_{k,m,n}\,(d\psi-d\phi)\,,
\label{eq:omkmnparts1}
\end{equation}
for some functions $\widetilde{\mu}_{k,m,n}$ and $\widetilde{\zeta}_{k,m,n}$.
The scalars, $ \widetilde{\mu}_{k,m,n}$ and $\widetilde{\zeta}_{k,m,n}$, can be solved separately by defining 
\begin{equation}
\widetilde{\mu}^S_{k,m,n} \equiv \widetilde{\mu}_{k,m,n} +\frac{R_y}{4\sqrt{2}}\frac{r^2+a^2\sin^2\theta}{\Sigma}\widetilde{\mathcal{F}}_{k,m,n}+\frac{R_y\,b_4^2}{4\sqrt{2}} \,\frac{\Delta_{2k,2m,2n}}{\Sigma}\,,
\end{equation}
Then $\widetilde{\mu}^S_{k,m,n}$ satisfies
\begin{equation}
\mathcal{L}\, \widetilde{\mu}^S_{k,m,n} = 
 \frac{R_y\,b_4^2}{\sqrt{2}}\frac{1}{(r^2+a^2)\cos^2\theta \, \Sigma} 
\left(\frac{(k-m)^2(k+n)^2}{k^2} \Delta_{2k,2m+2,2n}
+ \frac{(nm)^2}{k^2} \Delta_{2k,2m,2n-2}\right) ,
\end{equation} 
and $\widetilde{\zeta}_{k,m,n}$ is determined by ($s_{\theta}=\sin\theta$, $c_{\theta}=\cos\theta$)
\begin{equation}
\begin{aligned}
\partial_r \widetilde{\zeta}_{k,m,n} &\,=\, 
\frac{r^2 \cos2\theta-a^2 s_{\theta}^2}{r^2+a^2 s_{\theta}^2} \partial_r\widetilde{\mu}_{k,m,n}
 -\frac{r \sin2\theta}{r^2+a^2 s_{\theta}^2}\partial_\theta \widetilde{\mu}_{k,m,n}\\
&\quad+\frac{\sqrt{2}R_y\,r}{\Sigma(r^2+a^2 s^2_{\theta})}
\left[ b_4^2 \Bigl( m s^2_{\theta} + n c^2_{\theta} -\frac{mn}{k} \cos 2 \theta \Bigr) \Delta_{2k,2m,2n}-\frac{a^2 (2r^2+a^2)s^2_{\theta}c^2_{\theta}}{\Sigma}\widetilde{\mathcal{F}}_{k,m,n}\right] \,,\\
\partial_\theta\widetilde{\zeta}_{k,m,n}&\,=\,\frac{r(r^2+a^2) \sin2\theta}{r^2+a^2 s_{\theta}^2}\partial_r \widetilde{\mu}_{k,m,n}+ \frac{r^2 \cos2\theta-a^2 s_{\theta}^2}{r^2+a^2 s_{\theta}^2} \partial_\theta\widetilde{\mu}_{k,m,n}\\
&\quad+\frac{R_y\,\sin 2 \theta}{\sqrt{2}\,\Sigma\,(r^2+a^2 s^2_{\theta})}
\left[ b_4^2 \Bigl( - m r^2+ n (r^2+a^2) - \frac{mn}{k} (2r^2+a^2) \Bigr) \Delta_{2k,2m,2n} \phantom{\frac{r^2}{\Sigma}} \right. \\
&\qquad\qquad\qquad\qquad\qquad\qquad{}
 \left. +\frac{a^2 r^2 (r^2+a^2)\cos 2\theta}{\Sigma}\widetilde{\mathcal{F}}_{k,m,n}\right] .
\end{aligned} \label{eq:zeta-eqns}
\end{equation}

\noindent To solve the equations for $\widetilde{\mathcal{F}}_{k,m,n} $ and $\widetilde{\mu}^S_{k,m,n}$, we find the generating function $F_{2k,2m,2n}$ that solves the equation
\begin{align}
\cL F_{2k,2m,2n}={\Delta_{2k,2m,2n}\over (r^2+a^2)\cos^2\theta\,\, \Sigma} \;.
\end{align}
The solution to this problem is given by
\begin{equation} 
F_{2k,2m,2n}\,=\,-\!\sum^{j_1+j_2+j_3\le k+n-1}_{j_1,j_2,j_3=0}\!\!{j_1+j_2+j_3 \choose j_1,j_2,j_3}\frac{{k+n-j_1-j_2-j_3-1 \choose k-m-j_1,m-j_2-1,n-j_3}^2}{{k+n-1 \choose k-m,m-1,n}^2}\,
\frac{\Delta_{2(k-j_1-j_2-1),2(m-j_2-1),2(n-j_3)}}{4(k+n)^2(r^2+a^2)}\,,
\label{eq:geneF}
\end{equation}
where
\begin{equation} 
{j_1+j_2+j_3 \choose j_1,j_2,j_3}\equiv \frac{(j_1+j_2+j_3)!}{j_1! j_2! j_3!}\,.
\end{equation}
In terms of $F_{2k,2m,2n}$, the form of $\widetilde{\mathcal{F}}_ {k,m,n}$ and $\widetilde{\mu}_{k,m,n}$ for general $k,m,n$ is
\begin{align}
\widetilde{\mathcal{F}}_{k,m,n} &= 4\, b_4^2\biggl[\frac{m^2 (k+n)^2}{k^2}\,F_{2k,2m,2n}+\frac{n^2 (k-m)^2}{k^2}\,F_{2k,2m+2,2n-2}\biggr],
 \notag\\
\widetilde{\mu}_{k,m,n}&= \frac{R_y\,b_4^2}{\sqrt{2}}\,\biggl[ 
\frac{(k-m)^2(k+n)^2}{k^2} F_{2k,2m+2,2n}
+\frac{m^2 n^2}{k^2} F_{2k,2m,2n-2} -\frac{\Delta_{2k,2m,2n}}{4\,\Sigma}\biggr]
\label{cFmuOrig}\\
&\hspace{30ex}
-R_y\,\frac{r^2+a^2\,\sin^2\theta}{4\sqrt{2}\,\Sigma}\,\widetilde{\mathcal{F}}_{k,m,n}
+\frac{R_y\,\,b^2}{4\sqrt{2}\,\Sigma}\,.\notag
\end{align} 
In this expression for $\widetilde{\mathcal{F}}_{k,m,n}$ and $\widetilde{\mu}_{k,m,n}$ it should
be understood that, when the coefficient of one of the $F$ functions is zero, the
term is zero. The term proportional to $b$ in the last line of  (\ref{cFmuOrig}) is a harmonic piece that we can freely add to the solution of the Poisson equation for $\widetilde{\mu}^S_{k,m,n}$. Once $\widetilde{\mathcal{F}}_{k,m,n}$ and $\widetilde{\mu}_{k,m,n}$ are determined, $\widetilde{\zeta}_{k,m,n}$ can be found by integrating \eqref{eq:zeta-eqns} on a case-by-case basis. The coefficient $b$ will be fixed by regularity in the next subsection.

\subsection{Regularity}
\label{subsec:RegCondOrig}

Generic solutions to the BPS equations are not necessarily regular.   We have already removed much of the singular behavior by coiffuring the modes in the sources for the second layer.  However, there is still a final step in which harmonic solutions (zero-modes) are chosen so as to cancel any remaining singular behaviour.  Singularities typically occur at the supertube and where coordinates degenerate. This means one should start by examining $r=0$, $\theta=0$ and  $r=0$, $\theta=\pi/2$.   According to \cite{Bena:2015bea,Bena:2017xbt}, to remove such singularities one should require  $\widetilde{\mu}_{k,m,n}$ and $\widetilde{\zeta}_{k,m,n}$ to vanish at $r=0$, $\theta=0$.   This fixes the constant, $b$, in $\widetilde{\mu}_{k,m,n}$ to be
\begin{equation} \label{eq:RegOrig}
b ~=~ b_4 \, \left[{k \choose m} {k+n-1 \choose n}\right]^{-\frac{1}{2}}\,,
\end{equation}
and the same condition for $\widetilde{\zeta}_{k,m,n}$ fixes the integration constant of \eqref{eq:zeta-eqns}.   By examining the metric near the supertube locus one finds that it is non-singular provided that the following condition is satisfied:
\begin{equation}
\frac{Q_1 Q_5}{R_y^2} ~=~  a^2 ~+~ \frac{b^2}{2}\,,
\label{SSreg1}
\end{equation}
This is the supergravity dual of the CFT ``strand budget''
\begin{equation}
N_{++} ~+~ k \, N_{k,m,n,q=0} ~=~  N_1 N_5\,,
   \label{sbudget2}
\end{equation} 
where $N_{++} \sim a^2$ is the number of $|++\rangle_1$ that remain in the configuration. 

\subsection{Conserved charges}
\label{subsec:ConsChar}

One can check that all the superstratum solutions described above are asymptotic to AdS$_3\times$S$^3\times$T$^4$ (or $K3$). The conserved charges can be extracted from the metric from the large-distance behavior of the scalars $Z_1$, $Z_2$ and $Z_4$ and the one-forms $\beta$ and $\omega$. The fluctuating modes fall off much faster than the charge monopoles and so the D-brane charges $Q_1$ and $Q_5$, given by the asymptotic behavior of $Z_1$ and $Z_2$ do not change. 

One can compute the momentum charges from the large-distance behavior of $\cF$ and the left and right five-dimensional angular momenta of the solutions from the $d\phi d\psi$ component of the metric which can be obtained by looking at the $d\phi + d\psi$ legs of the one-form $\beta+\omega$. We have the generic expressions
\begin{equation}
\beta_\phi +\beta_\psi +\omega_\phi +\omega_\psi \sim \sqrt{2}\,\frac{J_R - J_L \cos 2\theta}{r^2}\,,\qquad \cF \sim-\frac{2 Q_P}{r^2}
\label{asmpmoms}
\end{equation} 
which gives
\begin{equation}
\label{eq:ConsChargesOrig}
J_R \,=\, \frac{R_y}{2} \left(a^2 +\frac{m}{ k} \,b^2 \right)\,,\qquad J_L \,=\, \frac{R_y}{2} \,a^2\,,\qquad Q_P = \frac{m+n}{2 k} \,b^2\,.
\end{equation} 

Finally, one can relate the conserved charges to their quantized values according to the string coupling, $g_s$, the Regge slope, $\alpha'$, and the volume of the four-dimensional manifold, $V_4$, using the following relations
\begin{equation}
 \label{eq:Relationquantizedcharges}
 \begin{aligned}
n_1 & ~=~  \frac{V_4}{(2\pi)^4\,g_s\,\alpha'^3}\,Q_1 \,, \quad    n_5  ~=~    \frac{1}{g_s\,\alpha'}\,Q_5 \,,\qquad n_p = \frac{V_4 R_y^2}{(2\pi)^4\,g_s^2\,\alpha'^4}\,Q_p   ~=~    \frac{R_y^2 \,n_1 n_5}{Q_1 Q_5}\, Q_p  \,, \\
j_{L, R} & ~=~  \frac{V_4\,R_y}{(2\pi)^4 g_s^2 \alpha'^4}\, J_{L, R} ~=~   \frac{R_y \,n_1 n_5}{Q_1 Q_5}\, J_{L, R}\,,
\end{aligned}
\end{equation}
where  $\cN$ is defined in (\ref{cNdefn}). This gives
\begin{equation}
j_R \,=\, \frac{\cN}{2}\left(a^2+\frac{m}{ k} b^2\right)\,,\qquad j_L \,=\, \frac{\cN}{2} a^2\,,\quad n_p \,=\, \frac{\cN}{2}\frac{m+n}{k} b^2\,.
\end{equation} 
These solutions are in the black-hole regime if the cosmic censorship bound is satisfied $n_1 n_5 n_p - j_L^2>0$, which is satisfied if 
\begin{equation}
\frac{b^2}{a^2} > \frac{k}{n+\sqrt{(k-m+n)(m+n)}}\,.
\end{equation} 

There are seven free parameters for the single-mode superstratum:  $Q_1, Q_5, k, m, n, a$ and $b$ and there are five bulk  charges: $Q_1, Q_5, Q_P,  j_{L}$ and $j_{R}$ as well as the regularity condition (\ref{SSreg1}), and so there can be, at most, one free ``rational'' parameter.  Indeed,  suppose $Q_1$ and $Q_5$ are fixed, and note that  $j_{L}$ fixes $a$.  The regularity condition (\ref{SSreg1}) then fixes $b$, which means that $j_{R}$ fixes $\frac{m}{k}$ and so $Q_P$ fixes $\frac{n}{k}$.  Thus, ignoring the issues of rational arithmetic, one can take $k$ to be the remaining free parameter.   The simplest classes of single-mode solutions that have been studied to date are the $(1,0,n,q=0)$ \cite{Bena:2016ypk,Bena:2017upb,Tyukov:2017uig,Bena:2018bbd, GreenSuperstrata}, $(2,1,n,q=0)$ \cite{Bena:2017geu,Bena:2017upb} and $(k,0,1,q=0)$ \cite{Bena:2018mpb} families.  Once the bulk charges are fixed, these families have no free parameters.  Part of our purpose here is to address this limitation:  to find relatively simple, explicitly-known families with free parameters.

\section{Supercharged single-mode Superstrata}
\label{sec:SuperchargedSuperstrata}

We now review the supergravity solutions for supercharged superstrata recently obtained in \cite{Ceplak:2018pws}.  Our discussion will closely parallel that of Section \ref{sec:OriginalSuperstrata}, but all the ``supercharged'' quantities  will be denoted with a `` $\widehat{\phantom{a}}${ }'' as opposed to a `` $\widetilde{\phantom{a}}${ }''' .

\subsection{The first layer}

The non-trivial excitations in the first layer of the BPS system now have the form:
\begin{equation}
\label{eq:SupTheta}
 \widehat{z}_{k,m,n} ~=~ 0\,, \qquad \widehat{\vartheta}_{k,m,n} =
\sqrt{2}\, \Delta_{k, m, n}\left[\,\frac{\Sigma}{r\sin\theta}\, \Omega^{(1)}\, \sin{v_{k,m,n}}+ \left(\Omega^{(2)} + \Omega^{(3)}\right)\cos{v_{k,m,n}}\,\right]\,,
\end{equation}
where $ \Delta_{k, m, n}$ and $v_{k,m,n}$ are defined in \eqref{Delta_v_kmn_def} and the self-dual basis, $\Omega^{(i)}$, $i=1,2,3$, is defined in \eqref{selfdualbasis}.  It is elementary to check that these modes satisfy:
\begin{equation}
 \cD\,\widehat{\vartheta}_{k,m,n} = 0, \qquad \widehat{\vartheta}_{k,m,n} \wedge d\beta=0,\qquad \widehat{\vartheta}_{k,m,n} = *_4 \widehat{\vartheta}_{k,m,n}\,.
 \label{schg1stlayer}
\end{equation}
This new fluctuating structure is remarkably simple.  Indeed, one should note the orthogonality in the middle equation of  (\ref{schg1stlayer}):
\begin{equation}
 \Theta_3 \wedge \widehat{\vartheta}_{k,m,n}  ~=~d\beta \wedge \widehat{\vartheta}_{k,m,n}  ~=~ 0\,,
\end{equation}
which allows one to take $\widehat{z}_{k,m,n} = 0$ in solving the first layer of the BPS equations.  Moreover, the exterior products $\widetilde{\vartheta}_{k,m,n}  \wedge \widehat{\vartheta}_{k,m,n}$ and  $\widehat{\vartheta}_{k,m,n}  \wedge \widehat{\vartheta}_{k,m,n}$ are ``self-coiffuring'' in that the oscillations cancel and the only residual part are the RMS values of the excitations.  This will be important in the ``hybrid superstrata,'' discussed in Section \ref{sec:Hybrids}. We also note that the expression for $ \widehat{\vartheta}_{k,m,n}$ is precisely the coefficient of $\frac{m n}{k}$ in  $\widetilde{\vartheta}_{k,m,n}$ (see (\ref{zthetatilde})).  This was important  for orthogonalizing and mixing the original and supercharged superstrata states in the dual CFT. For rather similar reasons, this observation will play an essential role in Section \ref{sec:multiSS}, where we show how to coiffure generic, multi-mode superstrata.

The field content of the first layer of the supercharged superstratum is\footnote{Compared to \cite{Ceplak:2018pws}, we have replaced $\{m+1,n+1\}\rightarrow\{m,n\}$ and we have added a $\sqrt{2}$ factor to make the comparison with the original superstratum solution on an equal footing.}:
\begin{equation}
 \label{eq:ZThetasinglemodeSupercharged}
 \begin{aligned}
 \widehat{Z}_1  & ~=~  \frac{Q_1}{\Sigma}\,,\qquad \widehat{\Theta}_2 ~=~  c_2\, \frac{R_y}{2\, Q_5}\, \widehat{\vartheta}_{2k,2m,2n} \,, \qquad   \widehat{Z}_2 ~=~\frac{Q_5}{\Sigma} \,,   \qquad  \widehat{\Theta}_1 ~=~ 0 \,, \\
 \widehat{Z}_4 & ~=~ 0 \,, \qquad\quad \widehat{\Theta}_4~=~ c_4\, \widehat{\vartheta}_{k,m,n} \,.
 \end{aligned}
\end{equation}
In analogy with the original superstrata, we have introduced a secondary mode in $\widehat{\Theta}_2$.  However, as we described above, the supercharged modes are ``self-coiffuring'' and so we will ultimately set $c_2 =0$. On the other hand, such a term will be essential once in the hybrid superstratum solutions.

\subsection{The second layer}
\label{sec:2ndLaySup}

As with the original superstrata, one expects the sources of the second layer of BPS equations to have an RMS part and an oscillating part that depends upon $v_{2k,2m,2n}$.  However, because $\widehat{Z}_4 = 0$ and $\widehat{\Theta}_4$ is ``self-coiffuring,'' there are no  terms that depend on $v_{2k,2m,2n}$ generated by  $\widehat{Z}_4$ and  $\widehat{\Theta}_4$.  Thus the only oscillating source is proportional to $c_2$ and so we take:
\begin{equation} 
c_2 ~=~ 0\,.
\label{Supcoiff}
\end{equation} 
The solution for $\omega$ and $\cF$ is given by the sums of the seed supertube solutions and the solution for the new pieces:
\begin{equation} 
\omega ~=~ \omega_0  ~+~ \widehat{\omega}_{k,m,n}\,,  \qquad \cF ~=~0 ~+~\widehat{\mathcal{F}}_{k,m,n} \,.
\label{omegaSup}
\end{equation} 
The equations \eqref{layer2eq} for $\widehat{\omega}_{k,m,n}$ and $\widehat{\mathcal{F}}_{k,m,n}$ reduce to:
\begin{align}
&d \widehat{\omega}_{k,m,n} + *_4 d \widehat{\omega}_{k,m,n} +\widehat{\mathcal{F}}_{k,m,n}\,d\beta
= 0  , \nonumber\\
& \mathcal{L} \,  \widehat{\mathcal{F}}_{k,m,n}
=
\frac{4\,c_{4}^2}{(r^2+a^2)\cos^2\theta \, \Sigma} \,
\left[ \, \Delta_{2k,2m,2n} \,+\,\Delta_{2k,2m+2,2n-2}\, \right] ,
\label{eq:F&omegaEqSup}
\end{align}

\noindent As in Section \ref{sec:2ndLayOrig}, one can solve directly the equation for $\widehat{\mathcal{F}}_{k,m,n}$ using the generating function $F_{k,m,n}$, \eqref{eq:geneF}. As for $\widehat{\omega}_{k,m,n} $, we define
\begin{equation}
\widehat{\omega}_{k,m,n} \;\equiv\;  \widehat{\mu}_{k,m,n} \,(d\psi+d\phi) + \widehat{\zeta}_{k,m,n}\,(d\psi-d\phi)\,.
\end{equation}
By shifting $ \widehat{\mu}_{k,m,n}$
\begin{equation}
\widehat{\mu}^S_{k,m,n} \equiv \widehat{\mu}_{k,m,n} +\frac{R_y}{4\sqrt{2}}\frac{r^2+a^2\sin^2\theta}{\Sigma}\widehat{\mathcal{F}}_{k,m,n}\,,
\end{equation}
we have
\begin{equation}
\mathcal{L}\, \widetilde{\mu}^S_{k,m,n} = 
 \frac{R_y\,c_4^2}{\sqrt{2}}\,\frac{ \Delta_{2k,2m+2,2n}
+\Delta_{2k,2m,2n-2}}{(r^2+a^2)\cos^2\theta \, \Sigma} .
\end{equation}
In terms of $F_{2k,2m,2n}$, \eqref{eq:geneF}, the form of $\widehat{\mathcal{F}}_ {k,m,n}$ and $\widehat{\mu}_{k,m,n}$ is 
\begin{align}
\widehat{\mathcal{F}}_{k,m,n} &= 4\, c_4^2\,\biggl[ \,F_{2k,2m,2n}+ \,F_{2k,2m+2,2n-2}\biggr],
\label{cFmuSup} \\
\widehat{\mu}_{k,m,n}&= \frac{R_y\,c_4^2}{\sqrt{2}}\,\biggl[ 
\, F_{2k,2m+2,2n} + F_{2k,2m,2n-2} ,\biggr] -R_y\,\frac{r^2+a^2\,\sin^2\theta}{4\sqrt{2}\,\Sigma}\,\widehat{\mathcal{F}}_{k,m,n}
+\frac{R_y\,\,c^2}{4\sqrt{2}\,\Sigma}\,,\nonumber
\end{align} 
where the term proportional to $c$ is a harmonic piece that we can add to the solution of the Poisson equation for $\widehat{\mu}^S_{k,m,n}$ and will be fixed by regularity in the next sub-section. Then, $\widetilde{\zeta}_{k,m,n}$ is determined by integrating ($s_{\theta}=\sin\theta$, $c_{\theta}=\cos\theta$)
\begin{equation}
\begin{aligned}
\partial_r \widehat{\zeta}_{k,m,n} &\,=\, 
\frac{r^2 \cos2\theta-a^2 s_{\theta}^2}{r^2+a^2 s_{\theta}^2} \partial_r\widehat{\mu}_{k,m,n}
 -\frac{r \sin2\theta}{r^2+a^2 s_{\theta}^2}\partial_\theta \widehat{\mu}_{k,m,n}-\frac{\sqrt{2}a^2 R_y\,r \,(2r^2+a^2)s^2_{\theta}c^2_{\theta}}{\Sigma^2(r^2+a^2 s^2_{\theta})}\widehat{\mathcal{F}}_{k,m,n}\,,\\
\partial_\theta\widehat{\zeta}_{k,m,n}&\,=\,\frac{r(r^2+a^2) \sin2\theta}{r^2+a^2 s_{\theta}^2}\partial_r \widehat{\mu}_{k,m,n}+ \frac{r^2 \cos2\theta-a^2 s_{\theta}^2}{r^2+a^2 s_{\theta}^2} \partial_\theta\widehat{\mu}_{k,m,n} \\
&\quad\,\,+\frac{R_y\,a^2 r^2 (r^2+a^2) \sin 2 \theta \cos 2\theta}{\sqrt{2}\,\Sigma^2\,(r^2+a^2 s^2_{\theta})}\widehat{\mathcal{F}}_{k,m,n} .
\end{aligned} \label{eq:zeta-eqnsSup}
\end{equation}
%

\subsection{Regularity and conserved charges}

The smoothness of supercharged solutions  closely follows that of the original superstrata and are detailed in Section \ref{subsec:RegCondOrig}. The scalar $\widehat{\mu}_{k,m,n}$ must vanish at ($r=0$, $\theta=0$) which requires that
\begin{equation}
\label{eq:RegSup}
c = c_4 \,\frac{k}{\sqrt{m n (k-m)(k+n)}} \,\left[{k \choose m} {k+n-1 \choose n}\right]^{-\frac{1}{2}}\,.
\end{equation}
Moreover, the absence of singularity at ($r=0$, $\theta=\pi/2$) relates the charges of the solution to the Fourier coefficients
\begin{equation}
\frac{Q_1 Q_5}{R_y^2} \,=\, a^2 + \frac{c ^2}{2}.
\label{SSreg2}
\end{equation}

As before, the conserved charges can be extracted from the metric from the large-distance behavior of the scalars $Z_1$, $Z_2$ and $\cF$ and the one-forms $\beta$ and $\omega$.  Indeed, from the supercharged analogue of (\ref{asmpmoms}) we can read off
\begin{equation}
J_R \,=\, \frac{R_y}{2} \left(a^2 +\frac{m}{ k} \,c^2 \right)\,,\qquad J_L \,=\, \frac{R_y}{2} \,a^2\,,\qquad Q_P = \frac{m+n}{2 k} \,c^2\,.
\end{equation}
The relations between the conserved charges and the Fourier coefficients of the supercharged superstrata, $a$ and $c$, are exactly the same as the ones for the original superstrata \eqref{eq:ConsChargesOrig} but with $b$ replaced by $c$.   The expressions for the quantized charges are given by \eqref{eq:QuantizedChargeOrig} (by setting $b=0$).

\section{Hybrid superstrata}
\label{sec:Hybrids}

As we described in the introduction, constructing multi-mode superstrata affords the possibility of making microstate geometries with additional moduli that represent the numbers, $N_{k,m,n,q}$, of excited strands with each kind of excitation.  In supergravity, these numbers are related,  via (\ref{NandFCnorms}), to the ``leading'' Fourier coefficients that appear in $\Theta_4$.   We will investigate generic multi-mode superstrata in Section \ref{sec:multiSS}.  Here our goal is more modest, but more completely executed: 
we will combine the original single-mode superstratum solution, detailed in Section \ref{sec:OriginalSuperstrata}, with the single-mode supercharged superstratum solution, detailed in Section \ref{sec:SuperchargedSuperstrata} to obtain a hybrid superstratum with an extra modulus obtained through the independent  Fourier coefficients. In Section \ref{sec:multiSS}, those hybrid superstrata will prove to be the crucial elementary components to construct multi-mode superstrata.

Combining two single-mode superstrata is straightforward for the first layer of BPS equations \eqref{layer1} thanks to linearity but it usually requires a great deal of effort to obtain the explicit solution to the second layer \eqref{layer2eq} because of the quadratic terms that  non-trivially mix the modes.  However, as we noted before, the supercharged modes and their original partners are  ``self-coiffuring'' in that  $\widetilde{\vartheta}_{k,m,n} \wedge \widehat{\vartheta}_{k,m,n}$,  is actually independent of $v$, $\phi$ and $\psi$.  This makes the computation, and the resulting solution, much simpler.    In this section, we will obtain explicit solutions for new hybrid single-mode superstrata with {\it eight} parameters: $Q_1, Q_5, k,m,n, a$ and the Fourier coefficients $b_4$ and $c_4$ of the superstratum modes.  The only constraint on smoothness will be a single equation generalizing (\ref{SSreg1})  and  (\ref{SSreg2}).  Thus there will be one constraint and five bulk ``state functions: $Q_1, Q_5, Q_P, J_L$ and $J_R$ this leaved two variables, which may be thought of as $k$ and the relative magnitude of $b_4$ and $c_4$.

\subsection{The first layer}

The first layer of BPS equations, (\ref{layer1}), is made of linear equations. At this level, pairing two BPS solutions simply consists in adding both solutions. Thus, our initial Ansatz for the field content is simply the sum of \eqref{eq:ZThetasinglemodeOrig} and \eqref{eq:ZThetasinglemodeSupercharged}
\begin{align}
 \label{eq:ZIAdSsinglemodeNew}
 \begin{aligned}
 Z_1 &~=~  \frac{Q_1}{\Sigma} + b_1 \,\frac{R_y}{2\, Q_5} \, \widetilde{z}_{2 k,2m,2n}\,, \qquad   \Theta_2 ~=~  b_1\, \frac{R_y}{2\, Q_5}\, \widetilde{\vartheta}_{2k,2m,2n} +c_2\, \frac{R_y}{2 \, Q_5}\, \widehat{\vartheta}_{2k,2m,2n} \,,  \\
 Z_2 & ~=~  \frac{Q_5}{\Sigma} \,,  \qquad  \Theta_1 ~=~ 0 \,, \qquad  Z_4  ~=~  b_4 \, \widetilde{z}_{k,m,n} \,, \qquad \Theta_4 ~=~    b_4\, \widetilde{\vartheta}_{k,m,n} + c_4\, \widehat{\vartheta}_{k,m,n} \,,
 \end{aligned}
\end{align}
where the original-superstratum contribution, $\widetilde{z}_{k,m,n}$ and $\widetilde{\vartheta}_{k,m,n}$, is given in \eqref{zthetatilde} and the supercharged-superstratum contribution,  $\widehat{\vartheta}_{k,m,n}$, is given in \eqref{eq:SupTheta}.

\subsection{The second layer}
\label{sec:2ndLayNew}

The quadratic terms in the second layer of BPS equations can induce oscillating parts that depend only upon $v_{2k,2m,2n}$, which one wants to cancel to ensure regularity. They come from the original superstratum, via $\widetilde{\vartheta}_{k,m,n}\wedge\widetilde{\vartheta}_{k,m,n}$, $\widetilde{\vartheta}_{2k,2m,2n}$, $\widetilde{z}_{k,m,n} \widetilde{\vartheta}_{k,m,n}$ and $\widetilde{z}_{2k,2m,2n} $, from the supercharged superstratum, via $\widehat{\vartheta}_{k,m,n}\wedge\widehat{\vartheta}_{k,m,n}$ and $\widehat{\vartheta}_{2k,2m,2n}$ and also from mixed terms, $\widetilde{\vartheta}_{k,m,n}\wedge\widehat{\vartheta}_{k,m,n}$ and $\widetilde{z}_{k,m,n} \widehat{\vartheta}_{k,m,n}$. As advertised in the preamble, the mixed term $\widetilde{\vartheta}_{k,m,n}\wedge\widehat{\vartheta}_{k,m,n}$ is intriguingly self-coiffuring. However, $\widetilde{z}_{k,m,n} \widehat{\vartheta}_{k,m,n}$ is not. This is why we have reinstated the secondary excitation mode with the Fourier coefficient $c_2$. Indeed, one can show that the Fourier coefficient of the oscillating sources vanish if
\begin{equation} 
b_1 ~=~b_4^2\,,\qquad c_2 ~=~ 2\,b_4\, c_4.
\label{Newcoiff}
\end{equation} 
One can now solve the second layer in a similar fashion as in Section \ref{sec:2ndLayOrig} and \ref{sec:2ndLaySup}. The solution for $\omega$ and $\cF$ is given by the sums of the  supertube solutions and the solution for the new pieces:
\begin{equation} 
\omega ~=~ \omega_0  ~+~ \omega_{k,m,n}\,,  \qquad \cF ~=~0 ~+~\cF_{k,m,n} \,.
\label{omegaNew}
\end{equation} 
The equations \eqref{layer2eq} for ${\omega}_{k,m,n}$ and ${\mathcal{F}}_{k,m,n}$ reduce to: 
\begin{align}
&d {\omega}_{k,m,n} + *_4 d {\omega}_{k,m,n} +{\mathcal{F}}_{k,m,n}\,d\beta
=
 \sqrt{2}\,b_{4} \,R_y\, \frac{\Delta_{2k,2m,2n}}{\Sigma}\left[ \left(\frac{m(k+n)}{k}\, b_4 - c_4\right) \,\Omega^{(2)} \right. \nonumber \\ & \hspace{8.8cm} \left. - \left( \frac{n(k-m)}{k}\, b_4+ c_4 \right) \,\Omega^{(3)} \right] , \nonumber\\
& \mathcal{L} \,  {\mathcal{F}}_{k,m,n}
= \frac{4}{(r^2+a^2)\cos^2\theta \, \Sigma} 
\left[ \left(\frac{m(k+n)}{k}\, b_4 - c_4\right)^{2} \Delta_{2k,2m,2n} \right. \label{eq:F&omegaEqNew} \\ 
& \hspace{5.1cm} \left.
+ \left( \frac{n(k-m)}{k}\, b_4+ c_4 \right)^{2} \Delta_{2k,2m+2,2n-2} \right] \nonumber,
\end{align}

\noindent We can straightforwardly check that taking $c_4 = 0$ or $b_4 =0$ in the foregoing equations reduces to the original-superstratum equations, \eqref{eq:F&omegaEqorig}, or the supercharged-superstratum equations, \eqref{eq:F&omegaEqSup}, respectively. Moreover, one can also note how non-trivial the pairing of the two solutions and its coiffuring \eqref{Newcoiff} are. For instance, if one looks to the coefficients in front of the functions $\Delta$ in the second equation in \eqref{eq:F&omegaEqNew}, one can see that the pairing does not simply consist in adding the contribution from the supercharged superstratum, which is $c_4^2$, with the contribution from the original superstratum, which is of the form $\gamma_{k,n,m}^2 b_4^2$ for some coefficient $\gamma_{k,n,m}$.  The source terms, rather remarkably, conspire to  complete the squares and lead to coefficients of the form $(\gamma_{k,n,m} b_4 \pm c_4)^2$. 

\noindent To solve the equations, we define, as before:
\begin{equation}
{\omega}_{k,m,n} \;\equiv\;  {\mu}_{k,m,n} \,(d\psi+d\phi) + {\zeta}_{k,m,n}\,(d\psi-d\phi)\,.
\label{eq:omkmnparts2}
\end{equation}
Then, by shifting $ {\mu}_{k,m,n}$
\begin{equation}
{\mu}^S_{k,m,n} \equiv {\mu}_{k,m,n} +\frac{R_y}{4\sqrt{2}}\frac{r^2+a^2\sin^2\theta}{\Sigma}{\mathcal{F}}_{k,m,n}+\frac{R_y\,b_4^2}{4\sqrt{2}} \,\frac{\Delta_{2k,2m,2n}}{\Sigma}\,,
\end{equation}
we have
\begin{equation}
\begin{split}
\mathcal{L}\, {\mu}^S_{k,m,n} = 
 \frac{R_y\,b_4^2}{\sqrt{2}} \frac{1}{(r^2+a^2)\cos^2\theta \, \Sigma} 
& \left[ \left(\frac{(k-m)(k+n)}{k} b_4 + c_4 \right)^2 \Delta_{2k,2m+2,2n}\right.
\\ &\left.  + \left(\frac{m n }{k} b_4 - c_4 \right)^2 \Delta_{2k,2m,2n-2}\right] .
\end{split}
\end{equation}
Then, ${\zeta}_{k,m,n}$ is determined by ($s_{\theta}=\sin\theta$, $c_{\theta}=\cos\theta$)
\begin{equation}
\begin{aligned}
\partial_r {\zeta}_{k,m,n} &\,=\, 
\frac{r^2 \cos2\theta-a^2 s_{\theta}^2}{r^2+a^2 s_{\theta}^2} \partial_r{\mu}_{k,m,n}
 -\frac{r \sin2\theta}{r^2+a^2 s_{\theta}^2}\partial_\theta {\mu}_{k,m,n}\\
&\quad+\frac{\sqrt{2}R_y\,r}{\Sigma(r^2+a^2 s^2_{\theta})}
\left[ b_4 \Bigl( \left(m s^2_{\theta} + n c^2_{\theta}\right) b_4 +\left(c_4 -\frac{mn}{k} b_4\right) \cos 2 \theta \Bigr) \Delta_{2k,2m,2n}\right. \\ 
& \hspace{3.4cm}\left. -\frac{a^2 (2r^2+a^2)s^2_{\theta}c^2_{\theta}}{\Sigma}{\mathcal{F}}_{k,m,n}\right] \,,\\
\partial_\theta{\zeta}_{k,m,n}&\,=\,\frac{r(r^2+a^2) \sin2\theta}{r^2+a^2 s_{\theta}^2}\partial_r {\mu}_{k,m,n}+ \frac{r^2 \cos2\theta-a^2 s_{\theta}^2}{r^2+a^2 s_{\theta}^2} \partial_\theta{\mu}_{k,m,n}\\
&\quad+\frac{R_y\,\sin 2 \theta}{\sqrt{2}\,\Sigma\,(r^2+a^2 s^2_{\theta})}
\left[ b_4 \Bigl( \left(- m r^2+ n (r^2+a^2)\right) b_4 + (2r^2+a^2)\left(c_4- \frac{mn}{k}b_4\right) \Bigr) \Delta_{2k,2m,2n} \phantom{\frac{r^2}{\Sigma}} \right. \\
&\hspace{4cm}
 \left. +\frac{a^2 r^2 (r^2+a^2)\cos 2\theta}{\Sigma}{\mathcal{F}}_{k,m,n}\right] .
\end{aligned} \label{eq:zeta-eqnsNew}
\end{equation}

\noindent In terms of $F_{2k,2m,2n}$, \eqref{eq:geneF},the form of ${\mathcal{F}}_ {k,m,n}$ and ${\mu}_{k,m,n}$ for general $k,m,n$ is
\begin{align}
{\mathcal{F}}_{k,m,n} =& 4 \,\biggl[ \left(\frac{m(k+n)}{k}\, b_4 - c_4\right)^2\,F_{2k,2m,2n}+ \left( \frac{n(k-m)}{k}\, b_4+ c_4 \right)^{2} \,F_{2k,2m+2,2n-2}\biggr],
 \notag\\
{\mu}_{k,m,n} =& \frac{R_y}{\sqrt{2}}\,\biggl[ 
\left(\frac{(k-m)(k+n)}{k} b_4 + c_4 \right)^2F_{2k,2m+2,2n}
+ \left(\frac{m n }{k} b_4 - c_4 \right)^2 F_{2k,2m,2n-2} -\frac{b_4^2 \,\Delta_{2k,2m,2n}}{4\,\Sigma}\biggr]
\notag\\
&
-R_y\,\frac{r^2+a^2\,\sin^2\theta}{4\sqrt{2}\,\Sigma}\,{\mathcal{F}}_{k,m,n}
+\frac{R_y\,\,B^2}{4\sqrt{2}\,\Sigma}\,.\label{cFmuNew}
\end{align} 
Then, ${\zeta}_{k,m,n}$ can be found by integrating \eqref{eq:zeta-eqns}. The coefficient $B$ will be fixed by regularity in the next subsection.

\subsection{Regularity and conserved charges}

As with the original superstratum solutions, the hybrid solutions must be smooth at the center of space and at the supertube locus. The conditions are essentially the same as the ones detailed in Section \ref{subsec:RegCondOrig}. The scalar $\widehat{\mu}_{k,m,n}$ must vanish at ($r=0$, $\theta=0$) which requires that
\begin{equation}
 \label{eq:RegNew}
B\,= \,\left(b_4^2 + \frac{k^2}{m n (k-m)(k+n)} c_4^2\right)^{\frac{1}{2}}\, \left[{k \choose m} {k+n-1 \choose n}\right]^{-\frac{1}{2}}\,=\, \sqrt{b^2+c^2}\,
\end{equation}
where $b$ is the individual contribution from the original superstratum part \eqref{eq:RegOrig} and $c$ is the individual contribution from the supercharged superstratum part \eqref{eq:RegSup}. Moreover, the absence of singularity at ($r=0$, $\theta=\pi/2$) relates the charges of the solution to the Fourier coefficients
\begin{equation}
\frac{Q_1 Q_5}{R_y^2} \,=\, a^2 + \frac{b^2+ c ^2}{2}.
\label{SGbudget1}
\end{equation}

From this expression one can easily see what the pairing of the two single-mode superstrata has produced. The resulting single-mode superstratum has the same integers number $k$, $m$ and $n$ and has combined the Fourier coefficients of the two initial solutions, $b$ and $c$, into one regularity constraint. Moreover, as we will now describe, the values of the conserved charges are given by the addition of the two individual contributions of the superstratum modes.  

As before, the  conserved charges can be extracted from the metric from the large-distance behavior of the scalars $Z_1$, $Z_2$ and $\cF$ and the one-forms $\beta$ and $\omega$.   We find
\begin{equation}
J_R \,=\, \frac{R_y}{2} \left(a^2 +\frac{m}{ k} \,\left( b^2+c^2\right) \right)\,,\qquad J_L \,=\, \frac{R_y}{2} \,a^2\,,\qquad Q_P = \frac{m+n}{2 k} \,\left( b^2+c^2\right)\,.
\end{equation}
As in Section \ref{subsec:ConsChar}, the quantized charges are given by:
\begin{equation}
j_R \,=\, \frac{\cN}{2}\left(a^2+\frac{m}{k}\,(b^2+c^2)\right)\,,\qquad j_L \,=\, \frac{\cN}{2} a^2\,,\quad n_p \,=\, \frac{\cN}{2}\frac{m+n}{k} (b^2+c^2)\,,
\label{eq:QuantizedChargeOrig}
\end{equation}
where $\cN$ is defined in \eqref{cNdefn}. All the conserved quantities depend on $b^2+c^2$ only. We define 
\begin{equation}
b \equiv B \cos \alpha\,, \qquad c \equiv B \sin \alpha
\label{eq:alphaB}
\end{equation}
where $ \alpha\in \left[0, 2 \pi \right] $ and $B\in \mathbb{R}_+$. All the asymptotic charges are independent of $\alpha$ which corresponds to an internal degree of freedom. However, $\alpha$ has a great impact on the IR geometry since it is the ``interpolation" parameter between the single-mode original superstratum state at $\alpha=0$ to a single-mode supercharged superstratum state at $\alpha=\frac{\pi}{2}$ with non-trivial pairing terms when $\alpha$ is in between. Thus, this new family of solutions allows one to study a phase space of superposition of two superstratum states by varying the phase $\alpha$ which have the same macroscopic mass, charges and angular momenta.

\newpage
\section{Generic multi-mode superstrata}
\label{sec:multiSS}

Here we consider the broad class of multi-mode hybrid superstrata.  That is, we will consider the hybrid superstrata of Section \ref{sec:Hybrids}, but now with a superposition of any set of mode excitations in the first layer.  While we will not present the complete solution, we will  examine a potentially dangerous obstruction to making smooth solutions in the second layer of BPS equations that arises when one restricts purely to the original superstrata.  We will also show how the supercharged modes play a crucial role in getting around this obstruction.

\subsection{The multi-mode problem}
\label{ss:multi-mode}
The second layer of BPS equations is:
\begin{equation}
\label{BPSlayer2}
\cD \omega + *_4 \cD \omega +\cF d\beta ~=~ {\cal S}_1  \,, \qquad *_4 \cD *_4\left(\dot{\omega}- \frac12 \cD \cF\right)  ~=~{\cal S}_2  \,,
\end{equation}
where the sources are defined by:
 \begin{align}
{\cal S}_1 & ~\equiv~   Z_1 \Theta_1 + Z_2 \Theta_2 - 2 Z_4 \Theta_4 \,, \label{BPSsource1} \\ 
{\cal S}_2 &~\equiv~ \partial_v^2(Z_1 Z_2 - Z_4^2)- (\dot{Z}_1\dot{Z_2}- (\dot{Z}_4)^2)- \frac12 *_4(\Theta_1\wedge \Theta_2 - \Theta_4 \wedge \Theta_4)\,.\label{BPSsource2} 
\end{align}
Since these sources are quadratic in the fundamental fields of the first BPS layer, the sources will only involve the interactions between pairs of modes.  It therefore suffices to solve this system for the most general {\it two-mode, hybrid superstratum} because one can construct the general mode solution through superpositions of all the two-mode source interactions.   

Recall the definitions of the first-layer modes, (\ref{zthetatilde}) and  (\ref{eq:SupTheta}), that underpin the original and supercharged superstrata:
\begin{align}
\widetilde{z}_{k,m,n} &=\,R_y \,\frac{\Delta_{k,m,n}}{\Sigma}\, \cos{v_{k,m,n}},\notag \\
 \widetilde{\vartheta}_{k,m,n}& ~\equiv~ -\sqrt{2}\,
\Delta_{k,m,n}
\biggl[\left((m+n)\,r\sin\theta +n\left({m\over k}-1\right){\Sigma\over r \sin\theta}  \right)\Omega^{(1)}\sin{v_{k,m,n}} \label{ztilde}\\
&\hspace{20ex}
 +\left(m\left({n\over k}+1\right)\Omega^{(2)} +\left({m\over k}-1\right)n\, \Omega^{(3)}\right) \cos{v_{k,m,n}} \biggr]\nonumber
\,,
\label{thetatilde}
\end{align}
and
\begin{equation}
\widehat{\vartheta}_{k,m,n} ~\equiv~
\sqrt{2}\, \Delta_{k, m, n}\left[\,\frac{\Sigma}{r\sin\theta}\, \Omega^{(1)}\, \sin{v_{k,m,n}}+ \left(\Omega^{(2)} + \Omega^{(3)}\right)\cos{v_{k,m,n}}\,\right]\,.
\label{thetahat}
\end{equation}
%
 %
%
\noindent The general two-mode, hybrid superstratum has a  first BPS layer of the form:
\begin{align}
Z_1 & ~=~  \frac{Q_1}{\Sigma} ~+~ \frac{R_y}{2\, Q_5} \,  \Big[b_1 \, \widetilde{z}_{2 k_1,2m_1,2n_1} ~+~ b_2 \, \widetilde{z}_{2 k_2,2m_2,2n_2} ~+~ b_3 \,\widetilde{z}_{k_1+k_2,m_1+m_2,n_1+n_2} \Big]  \,,   \notag\\
Z_2 & ~=~ \frac{Q_5}{\Sigma} \,,  \qquad
Z_4  ~=~   b_4 \, \widetilde{z}_{k_1, m_1, n_1} ~+~ b_5 \, \widetilde{z}_{k_2, m_2, n_2} \,,\notag \\
\Theta_1  & ~=~  0 \,, \quad  \qquad \Theta_4   ~=~ b_4\, \widetilde{\vartheta}_{k_1, m_1, n_1} ~+~ b_5\,  \widetilde{\vartheta}_{ k_2, m_2, n_2} ~+~ c_4\, \widehat{\vartheta}_{ k_1, m_1, n_1} ~+~ c_5\,\widehat{\vartheta}_{k_2, m_2, n_2} \,,   \label{eq:Zansatz}\\
\qquad \Theta_2 & ~=~\frac{R_y}{2\, Q_5}\, \Big[ b_1\, \widetilde{\vartheta}_{2 k_1,2m_1,2n_1} ~+~ b_2\,  \widetilde{\vartheta}_{2 k_2,2m_2,2n_2} ~+~ b_3\,  \widetilde{\vartheta}_{k_1+k_2,m_1+m_2,n_1+n_2}  \notag\\ 
&\qquad \qquad ~+~ c_1\,\widehat{\vartheta}_{2 k_1,2m_1,2n_1} ~+~ c_2\, \widehat{\vartheta}_{2 k_2,2m_2,2n_2} ~+~ c_3\,\widehat{\vartheta}_{k_1+k_2,m_1+m_2,n_1+n_2}    \Big]\,, \notag
\end{align}
%

\subsection{The high-frequency sources in the second BPS layer}
\label{ss:modes}

The sources, (\ref{BPSsource1}) and (\ref{BPSsource2}), contain terms that are products of pairs of oscillating terms.  Using  elementary identities, one can rewrite these products in terms of circular functions that involve  sums and differences of the mode numbers.    We will define the ``high-frequency'' and  ``low-frequency'' sources to be those that result from the addition or, respectively, subtraction of mode numbers.   Since one has $k_1,k_2,m_1,m_2,n_1, n_2 \ge 0$,  the ``high-frequency'' sources do indeed have larger mode numbers than their ``low-frequency'' counterparts.  We will also define the ``high-frequency" sources to include the terms that come from the product of a non-fluctuating term (zero mode-number) with a  fluctuating mode.  

From a great deal of experience  \cite{Bena:2015bea,Bena:2016agb,Bena:2016ypk,Bena:2017geu,Bena:2017upb,Bena:2017xbt} we know that the BPS equations for the non-oscillating sources and the ``low-frequency'' sources have non-singular solutions that lead to smooth superstrata.  On the other hand, the  ``high-frequency'' sources generically lead to singular solutions of the BPS equations.  We should stress that this statement is not a theorem but reflects the results of many computations in which homogeneous solutions have been found to cancel putative singularities in ``low-frequency'' solutions but no such homogeneous solutions have been found for ``high-frequency'' solutions.  

In supergravity, the technique of ``coiffuring'' \cite{Bena:2013ora,Bena:2014rea,Bena:2015bea,Bena:2017xbt} was introduced to address this issue.  The idea is that ``high-frequency'' sources that lead to singularities can be combined in such a manner  as to cancel any singularity.  In simpler examples, like  asymptotically-AdS superstrata, this means cancelling all of the ``high-frequency'' sources.  In more complicated examples, like  asymptotically-flat superstrata, the high-frequency sources are not completely cancelled but are combined so as to remove singularities in the solutions. 

Since we are considering asymptotically-AdS superstrata, our goal here is to isolate the ``high-frequency'' sources, $\bar{\cal S}_j$, and cancel them completely. It is relatively easy to compute these sources from (\ref{BPSsource1}) and  (\ref{BPSsource2}), and we find:
\begin{align}
\bar {\cal S}_1~=~ \frac{R_y}{2\, \Sigma}\, &\bigg[(b_1 - b_4^2)\,  \widetilde{\vartheta}_{2 k_1,2m_1,2n_1}  ~+~ (b_2 - b_5^2)\,    \widetilde{\vartheta}_{2 k_2,2m_2,2n_2} ~+~ \,  (b_3 - 2\,b_4\, b_5)\,  \widetilde{\vartheta}_{k_1+k_2,m_1+m_2,n_1+n_2}   \nonumber  \\ & ~+~ ( c_1- 2\, b_4 \,c_4 )\,\widehat{\vartheta}_{2 k_1,2m_1,2n_1}   ~+~ ( c_2- 2\, b_5 \,c_5 )\,\widehat{\vartheta}_{2 k_2,2m_2,2n_2}   \nonumber \\ 
&~+~   \bigg( c_3  - 2\, (b_4 \,c_5+ b_5 \,c_4)  \nonumber \\ 
& \qquad \qquad~+~ \frac{2\,b_4 \,b_5}{k_1 k_2 (k_1+ k_2)}  (k_1 n_2 - k_2 n_1)(k_1 m_2 - k_2 m_1)\, \bigg)\,\widehat{\vartheta}_{k_1+k_2,m_1+m_2,n_1+n_2}\bigg]  \,, \label{BPS-hfsource1} 
\end{align}
and 
\begin{align}
\bar  {\cal S}_2 ~=~  \frac{4}{R_y\, \Sigma}\, \Big[& (m_1+n_1)^2 (b_4^2 -  b_1)\, \widetilde{z}_{2 k_1,2m_1,2n_1}  ~+~ (m_2+n_2)^2 (b_5^2 - b_2)\, \widetilde{z}_{2 k_2,2m_2,2n_2}  \nonumber  \\ & ~+~ \frac{1}{4}\, (m_1+ m_2 +n_1+n_2)^2 (2\,b_4\, b_5 -  b_3)\, \widetilde{z}_{k_1+k_2,m_1+m_2,n_1+n_2}  \,\Big] \,. \label{BPS-hfsource2} 
\end{align}
\noindent The obvious coiffuring identities are 
\begin{align}
b_1 & ~=~ b_4^2 \,, \qquad   b_2 ~=~ b_5^2 \,, \qquad   b_3 ~=~2 \, b_4 \, b_5\,,  \qquad c_1  ~=~ 2\, b_4 \,c_4 \,, \qquad   c_2 ~=~2\, b_5 \,c_5\,,\nonumber \\
 \qquad   c_3 &~=~2\,( b_4 \,c_5+ b_5 \,c_4)   ~-~ \frac{2\,b_4 \,b_5}{k_1 k_2 (k_1+ k_2)}  (k_1 n_2 - k_2 n_1)(k_1 m_2 - k_2 m_1)    \,, \label{coiff1} 
\end{align}
which means that all the ``high-frequency" sources do indeed vanish. 

There are several important things to note at this point. First and foremost, there are still four free parameters in these solutions: $b_4, b_5, c_4$ and $c_5$, which correspond to the fundamental, or ``leading,'' modes in $\Theta_4$ and thus, at linear order, to the four independent sets of CFT states.   Next we note that  the  expression for $c_3$, (\ref{coiff1}), has a  term proportional to $b_4 b_5 \sim b_3$.  There is therefore no solution with all of the $c_j =0$:  the supercharged modes are essential to a non-singular, coiffured solution.   

Put differently, there is no way to coiffure a generic set of two-mode superstrata using  the original superstratum modes {\it only}.  The dangerous source term is the one proportional to $b_4  b_5\, \widehat{\vartheta}_{k_1+k_2,m_1+m_2,n_1+n_2}$ in the last line of (\ref{BPS-hfsource1}).  This term comes from the fact the $b_3$ term produces a contribution to  $\bar {\cal S}_1$ that is proportional to $\frac{(m_1 +m_2)(n_1 +n_2)}{k_1 + k_2}\widehat{\vartheta}$ whereas the corresponding $b_4 b_5$ term is proportional to $(\frac{m_1 n_1}{k_1} +\frac{m_2 n_2}{k_2})\widehat{\vartheta}$. There is thus an imperfect cancellation from setting $b_3 ~=~ b_4 \, b_5$ and the residual part is proportional to $ (k_1 n_2 - k_2 n_1)(k_1 m_2 - k_2 m_1) \widehat{\vartheta}$.  This term is only non-trivial if neither of $(m_1,m_2)$ and $(n_1,n_2)$ are parallel to $(k_1,k_2)$.  Such generic two-mode solutions of the original superstratum have not been investigated because they are complicated and there was still this unresolved issue with coiffuring.   As we have seen, the supercharged modes provide an elegant resolution of this issue with {\it no additional constraints on the fundamental, physical Fourier coefficients.}

More broadly, there is now no further obstruction, in principal, to the construction of superstrata with an arbitrary number of excited modes.  As we noted above, the BPS equations have sources that are quadratic in the fundamental modes and so a generic multi-mode solution will simply result in a superposition of many two-mode interactions.  Thus, in addressing  the most general two-mode problem one has necessarily captured all the pieces of a generic multi-mode solution.  The analysis above therefore shows that, once one includes the supercharged modes, there are no longer any singular ``high frequency'' interactions that cannot be coiffured away easily. 

If one were to restrict to the original ($q=0$) superstrata alone then the most obvious way to remove the dangerous high frequency sources is to require  
\begin{equation}
(k_1 n_2 - k_2 n_1)(k_1 m_2 - k_2 m_1)   ~=~ 0\,,
\label{detcond}
\end{equation}
which restricts the modes to two-dimensional subspaces.  Such superstrata are thus restricted to intrinsically two-dimensional fluctuations and are consequently restricted in their ability to cover the dual CFT states. In the hybrid superstrata, we now have unconstrained modes with two families of unconstrained Fourier coefficients (the $b$'s and the $c$'s) which means that the superstratum fluctuations are described by two generic functions of three variables and can faithfully describe all the coherent CFT states (\ref{multistates}).

As we described in the introduction, the coiffuring constraints are also testable by computing three-point functions in the CFT \cite{Giusto:2015dfa}.   The secondary Fourier coefficients are $b_i$ and $c_i$, $i =1,2,3$ and the determination of  their values in terms of  $b_4, b_5, c_4$ and $c_5$ represents the prediction of  vevs.  The values of $b_i$ are perhaps less interesting as these are naturally produced in solution generating methods.  However, the values of the $c_i$, and especially the contribution of  $b_4$ and $b_5$ to $c_3$ should yield a very non-trivial test of holography.

\section{The ($2,1,n$) example}
\label{sec:example}

In this section, we illustrate the new type of superstratum solutions with a detailed example: the  ($2,1,n$) hybrid.  This is, in fact, the first non-trivial family since one must have $k \geq m+1\geq 2$ and $n\geq 1$.  From here on, we will reparametrize the Fourier coefficients, $b$ and $c$, with $\alpha$ and $B$ \eqref{eq:alphaB}  so as to make the internal degree of freedom manifest. 

\noindent The modes and the functions in (\ref{Delta_v_kmn_def}) are now simply 
\begin{equation}
v_{2,1,n} \,=\, (n+1)\frac{\sqrt{2} v}{R_y} +\phi-\psi\,, \qquad  \Delta_{2,1,n} \,=\, (1-\Gamma)\,\Gamma^{\frac{n}{2}} \,\cos \theta \,\sin\theta \,,
\end{equation}
where we have defined
\begin{equation}
\Gamma \,\equiv\, \frac{r^2}{a^2+r^2}.
\end{equation}
%
\subsection{The metric}
The scalar functions $\cP_{2,1,n}$, $\cF_{2,1,n}$ are given by: 
\begin{equation}
\begin{split}
\cP_{2,1,n} &\,=\, \frac{R_y^2}{\Sigma^2}\, \left( a^2 +\frac{B^2}{2} -(n+1) \,B^2\,\cos^2\alpha  \, (1-\Gamma)^2\,\Gamma^{n} \,\cos^2 \theta \,\sin^2\theta\right)\,, \\
\cF_{2,1,n} &\,=\, \frac{n(n+1)\,B^2}{2a^2(1-\Gamma)}  \left[ n \left(\cos\alpha+ \sqrt{\frac{n+2}{n}}\,\sin\alpha\right)^2 \left( G_n^{(1)}(\Gamma) + G_n^{(2)}(\Gamma) \cos^2 \theta\right)\right. \\
   & \hspace{2.3cm} \left. + (n+2) \left(\sin \alpha - \sqrt{\frac{n+2}{n}}\cos\alpha\right)^2 \left( G_{n+1}^{(1)}(\Gamma) + G_{n+1}^{(2)}(\Gamma) \sin^2 \theta\right) \right]\,,
\end{split}
\label{eq:P&F21n}
\end{equation}
where  
\begin{equation}
\begin{split} 
G_n^{(1)}(\Gamma) &\,\equiv\, \frac{\Gamma\, (1-\Gamma^n) -n (1-\Gamma)}{n^2(1+n)^2} \, ,\\
G_n^{(2)}(\Gamma) &\,\equiv\, \frac{-1-\Gamma+(1+n)^2 \,\Gamma^n +(1-2n(n+1))\,\Gamma^{n+1} +n^2\, \Gamma^{n+2}}{n^2(1+n)^2} \,,
\end{split}
\end{equation}
The angular momentum one-form, $\omega_{2,1,n}$, is given by:
\begin{equation}
\omega_{2,1,n} \,=\, \omega_0 + B^2\,\left[ \,\cos^2\alpha \,\,\widetilde{\omega}_{2,1,n} + \frac{\sin^2\alpha}{n(n+2)} \,\, \widehat{\omega}_{2,1,n} +\frac{\sin \alpha\, \cos\alpha}{\sqrt{n(n+2)}} \,\,\omega_{2,1,n}^{\text{mix}}\,\right]\,,
\label{eq:omega21n}
\end{equation}
where $\omega_0$ is the supertube contribution, (\ref{eq:stube2}),  while $\widetilde{\omega}_{2,1,n}$ and $ \widehat{\omega}_{2,1,n}$ are, respectively, the individual contributions from the original superstratum and the supercharged superstratum.  There is also a mixing, $\omega_{2,1,n}^{\text{mix}}$, coming from the interactions between the original and supercharged modes.  These are given by:
\begin{alignat}{1}
\widetilde{\omega}_{2,1,n} &\,=\, \frac{R_y}{2\sqrt{2}\,\Sigma} \biggl[\, \Gamma^{n+1} \left(n+2-(n+1)\,\Gamma - (n+1)\, (1-\Gamma )\,  \cos 2 \theta\right)\, \cos ^2\theta \,d\psi \notag \\
& \hspace{1.9cm} +
   \left(2 - (n \,( 1-\Gamma )+1)\, \Gamma ^n -(n+1)\, (1-\Gamma )\,\Gamma ^n  \cos 2 \theta  \right) \, \sin ^2 \theta \, d\phi \, \biggr]\,,\notag \\
\widehat{\omega}_{2,1,n} &\,=\, \frac{R_y}{2\sqrt{2}\,(1-\Gamma)^2\,\Sigma} \biggl[ \,\Gamma \left(4- (2 + n\,(1- \Gamma) )^2\Gamma ^n\right)\, \left( \cos ^2\theta \,d\psi  - \sin ^2 \theta \, d\phi \right)\\
& \hspace{3.3cm}+ 2 \, n (n+2)\, (\Gamma -1)^2 \, \sin ^2 \theta \, d\phi \,   \biggr]\,,\notag\\
\omega_{2,1,n}^{\text{mix}}& \,=\, \frac{R_y}{4\sqrt{2}\,(1-\Gamma)\,\Sigma} \,\left(2 - (n+1)(n+2)\, \Gamma^n +
 2 n (2 + n) \,\Gamma^{1 + n} - 
 n (1 + n) \,\Gamma^{2 + n} \right)\,\notag \\
 & \hspace{3.1cm} \times \, \sin ^2 2 \theta \,\,  \left(\, \Gamma \, d\psi + d\phi \, \right)\,. \notag
\end{alignat}
One can easily check that if one takes $\alpha = \frac{\pi}{2}$, then one recovers the fields of the ($2,1,n,q=1$) supercharged superstratum \cite{Ceplak:2018pws}\footnote{We remind that in our notation the ($2,1,n,q=1$) supercharged superstratum corresponds to the ($2,0,n-1,q=1$) solution in \cite{Ceplak:2018pws}.} whereas $\alpha=0$ gives the fields of the ($2,1,n,q=0$) original superstratum first constructed in \cite{Bena:2017upb}. Taking any value in between gives a non-trivial hybrid of those two solutions.

\subsection{Limiting geometries}

We now consider the interesting regime of parameters in which $a^2$  is much smaller than the underlying charges:
\begin{equation}
a^2 \ll Q_{1,5,P} \quad \Longrightarrow \quad a^2 \ll B^2\, .
\end{equation}
In this regime, one has an AdS$_3$ at large $r$ ($r \gg \sqrt{Q_P}$), which then transitions to a long extremal BTZ throat  (for $\sqrt{n}\, a \ll r  \ll  \sqrt{Q_P}$), which ultimately ends with a smooth cap ($r \ll \sqrt{n}\,  a$). We will essentially focus in the IR and UV limit of the metric. In the previous section we  gave all the pieces of the metric and if one combines them all into (\ref{sixmet}), it produces a very complicated and rather unedifying mess that we will not write explicitly.   However, we will look at some physically interesting regions of the geometry in which the metric simplifies significantly.  In particular, as observed in \cite{Bena:2018bbd, GreenSuperstrata}, an important part of the physics of the solutions is contained in the  geometry of the cap.

\subsubsection{The asymptotic geometry}

We start with the simplest limit:   $r\gg\sqrt{n}\, a$, in which the metric reduces to:
\begin{equation}
\begin{aligned}
 ds_\text{asym}^2 ~=~   \sqrt{Q_1 Q_5} \,  \bigg[&  \;\! \frac{dr^2}{r^2} \,-\, \frac{4\, r^2}{(2 a^2+ B^2)\,R_y^2}\, du \, dv  \,+\,  \frac{2n+1}{2 \, R_y^2}\, dv^2  \;\!  \\
  & + \, d\theta^2  \,+\,  \sin^2\theta \, \left(d\phi -\frac{dv}{\sqrt{2} \,R_y}\right)^2 \,+\, \cos^2 \theta \, \left(d\psi - \frac{dv}{\sqrt{2} \,R_y} \right)^2  \;\! \bigg] \,.
\end{aligned}
\label{eq:BTZmetric21n}
\end{equation}
This is simply a trivial $S^3$ fibration over a red-shifted extremal BTZ geometry. The left and right temperatures are 
\begin{equation}
T_L \,=\, \frac{\sqrt{n+\coeff{1}{2}}}{2\pi\, R_y}\, \qquad T_R \,=\,0\,.
\end{equation}
Note that the BTZ region is independent of the parameter $\alpha$.  This parameter represents an internal degree of freedom and does not change the macroscopic charges, mass and angular momenta of the solutions.

\subsubsection{The cap geometry}

The cap geometry is obtained by taking the limit $r\ll \sqrt{n} a$. We decompose the six-dimensional cap metric according to the free parameter $\alpha$ as follows
\begin{equation}
\begin{aligned}
 ds_\text{cap}^2 ~=~   \sqrt{Q_1 Q_5} \,  \bigg[&  \;\! \frac{dr^2}{r^2 + a^2} - \frac{2 \, a^2 (r^2 + a^2) }{(2 a^2+B^2)^2 \,R_y^2} \, (du+dv)^2  + \frac{2 a^2\,r^2}{(2 a^2+B^2)^2 \,R_y^2}\left(du -\left(1+\frac{B^2}{a^2} \right)  dv \right)^2 \;\!  \\
  & + \, d\theta^2  \,+\,  \cos^2 \alpha \,\, {d\widetilde{\Omega}_2}^2 \,+\,\sin^2\alpha\,\, {d\widehat{\Omega}_2}^2 \,+\,\sin \alpha\, \cos\alpha\,\, {d\Omega^\text{mix}_2}^2  \;\! \bigg] 
\,.
\end{aligned}
\label{eq:capmet}
\end{equation}
where
\begin{alignat}{1}
{d\widetilde{\Omega}_2}^2 &\,=\, \sin^2\theta \, \left(d\phi -\frac{\sqrt{2}\,a^2}{(2\,a^2 +B^2) \,R_y} \left(du+dv\right) \right)^2 + \cos^2 \theta \, \left(d\psi + \frac{\sqrt{2}\,a^2 \left(du-dv\right) - \sqrt{2}\,B^2 dv}{(2\,a^2 +B^2) \,R_y}  \right)^2,\notag\\
{d\widehat{\Omega}_2}^2 &\,=\, \sin^2 \theta \, \left(d\phi -\frac{\sqrt{2}\,a^2}{(2\,a^2 +B^2) \,R_y} \left(du+dv\right) + \frac{2\sqrt{2} \,B^2\, r^4}{n(n+2)\,a^4\,(2 a^2 + B^2) R_y}\, dv \right)^2 \\
&  \hspace{0.5cm}  + \,\cos^2 \theta \, \left(d\psi + \frac{\sqrt{2}\,a^2 \left(du-dv\right) - \sqrt{2}\,B^2 dv}{(2\,a^2 +B^2) \,R_y} -  \frac{2\sqrt{2} \,B^2\, r^4}{n(n+2)\,a^4\,(2 a^2 + B^2) R_y}\, dv   \right)^2\,, \notag\\
 {d\Omega^\text{mix}_2}^2 &\,=\,\frac{2 B^2}{(2 a^2+ B^2)\sqrt{n(n+2)}} \left[ \,- \sin^4 \theta\,  \left(d\phi  + \frac{\sqrt{2}\, r^2}{a^2 R_y} \,dv \right)^2 \,+\, \cos^4 \theta\,  \left(d\psi  - \frac{\sqrt{2}\, (a^2+ r^2)}{a^2 R_y} \,dv \right)^2 \right]\,.\notag
\label{eq:metriccontrSup}
\end{alignat}
The first line of (\ref{eq:capmet}) defines a (hugely red-shifted and boosted) global  AdS$_3$.   Indeed, the red-shift in front of the time coordinate is  given by:
\begin{equation}
\bigg(1+\frac{B^2}{2 a^2}\bigg)^{-1} ~=~ \frac{2 \,j_L}{N_1 N_5} \,.
\label{eq:redshitedt}
\end{equation}
The various of $d{\Omega}_2$ terms in (\ref{eq:capmet}) give the metric on the $U(1) \times U(1)$ defined by $(\phi,\psi)$. The $\phi$-circles and $\psi$-circles universally pinch-off  at $\theta=0$ and $\theta=\frac{\pi}{2}$, respectively and so the $(d\theta , d\phi, d\psi)$ components  describe a squashed $S^3$.  

It is remarkable to note that all the off-diagonal components of the metric that mix  $(d\phi, d\psi)$ with $(du, dv)$ are independent of $\theta$, and are either constant or depend solely upon $r$.   This means that they can be reduced to non-trivial Kaluza-Klein massless electromagnetic fields on the three-dimensional base defined by the AdS$_3$.   Thus, the dynamics of the six-dimensional cap metric excitations reduces to dynamics of the metric and massless vector fields on the three-dimensional base space described by ($r,u,v$).    For $\alpha=0$, the metric of the 3-sphere is reduces to that of the ($2,1,n,q=0$) original superstratum, $d\theta^2 + {d\widetilde{\Omega}_2}$. The $S^3$ is round and the Kaluza-Klein vector fields on the $S^3$  are constant and so the fibration is trivial. At $\alpha = \frac{\pi}{2}$  the solution is purely the ($2,1,n,q=1$) supercharged superstratum.  The  $S^3$ is also round but now the fibration is non-trivial in that the vector fields depend on $r$.   When $\alpha$ takes any other value, the mixed term, ${d\Omega^\text{mix}_2}^2$, warps the 3-sphere  with, once again, non-trivial $r$-dependent vector fields. 

The fact that the fibration of the cap metric has the simple Kaluza structure suggests that there may be some interesting consistent truncation of the six-dimensional physics of the cap to a gauged supergravity in three-dimensions.  The Kaluza-Klein fields are all of the correct form while the squashing may reflect the role of a Kaluza-Klein scalar in three dimensions.  Similar observations were made for complete metric of some of the original superstrata constructed in \cite{Bena:2017upb}.

\subsection{Scalar wave perturbations}

The scalar perturbations of the original  ($1,0,n,q=0$)  superstrata  have been studied fairly thoroughly  \cite{Bena:2017upb,Bena:2018bbd,GreenSuperstrata} because the massless wave equation is separable.  The highly coherent structure of such single-mode superstrata leads to results that differ significantly from what one would  expect  for a typical black-hole microstate. First, the spectrum of normalizable modes is linear with a mass gap and an energy gap between normalizable modes of order $4j_L/(n_1 n_5)$.  Secondly, the response function \cite{GreenSuperstrata}, while exhibiting some BTZ-like behaviour for short times, has, at large time scales, sharp, resonant echoes coming from the bound states in the cap.  All of this is hardly surprising because the original single-mode superstrata have no remaining degrees of freedom after the macroscopic charges are fixed.  Individually, they are   rigid, atypical, highly coherent states.

Much richer spectral structure and response functions should be obtained by studying scalar wave perturbation in multi-mode superstrata.  However, their complicated structure makes such an analysis very difficult to perform.  Our hybrid superstratum solutions offer something of an intermediate option. They have a similar structure to the one of the original single-mode superstrata but they have an internal degree-of-freedom given by $\alpha$. The purpose of this section is to illustrate, in a relatively friendly family of hybrid superstrata, the $\alpha$-dependence of the spectrum of normalizable modes.

We consider a minimally-coupled massless scalar perturbation on the top of our geometry
\begin{equation}
\frac{1}{\sqrt{-\det g}} \: \partial_M \left( \sqrt{-\det g}\: g^{MN} \partial_N \, \Phi \right) \:=\: 0\,,
\end{equation}
where $g_{MN}$ is the complete\footnote{We consider the overall geometry given in \eqref{sixmet} determined by $\cP_{2,1,n}$, $\cF_{2,1,n}$ and $\omega_{2,1,n}$ in \eqref{eq:P&F21n} and \eqref{eq:omega21n} and not only the cap metric.} six-dimensional metric. Such a scalar perturbation can be obtained by a small deformation of the transverse-$\IT^4$ metric \cite{Bombini:2017sge}.   Since the geometry is independent of $u$, $v$, $\phi$, and $\psi$, we can at least decompose into Fourier modes along these directions: 
\begin{equation}
\Phi = H(r,\theta)\,e^{i\left( \frac{\sqrt{2}\,\omega}{R_y} \,u\,+\,\frac{\sqrt{2}\, p}{R_y} \, v \,+\,q_1 \phi \,+\, q_2 \psi \right)}\,.
\label{eq:modeprofile}
\end{equation}
The wave equation gives
\begin{equation}
\bar{\cL}\, H(r,\theta) - V_r (r) \, H(r,\theta)  - V_\theta (\theta) \, H(r,\theta)\, + \left(\frac{r^2}{r^2+a^2} \right)^n \cW_n (r,\theta) \, H(r,\theta)\,=\, 0,
\label{eq:WEgenalpha}
\end{equation}
where we have defined the operator, $\bar{\cL}$, via
\begin{equation}
\bar{\cL} \,\equiv\, \frac{1}{r}\, \partial_r \big( r (r^2 + a^2) \, \partial_r  \big)  +  \frac{1}{\sin \theta \cos \theta}\partial_\theta \big( \sin \theta \cos \theta\, \partial_\theta   \big) .
\end{equation}
The angular and radial potentials, $V_r (r)$ and $V_\theta (\theta)$, are given by:
\begin{equation}
\begin{split}
V_\theta (\theta) \,\equiv \, &\frac{q_1^2}{\sin^2 \theta} \,+\, \frac{q_2^2}{\cos^2 \theta} \,-\, \frac{\sin\alpha \cos\alpha}{\sqrt{n(n+2)}}\,\frac{2 B^2 \,\omega}{a^4}\, \cos 2\theta \,\left(B^2 \omega+ a^2 (q_1 -q_2 +2\,\omega)\right) \\
& \,-\, \frac{\sin^2\alpha \cos^2\alpha}{n(n+2)} \frac{B^4 \,\omega^2}{2 a^4} \,\cos 4\theta\,, \\
V_r(r) \,\equiv\, &\frac{a^2 (q_2+p-\omega)^2}{r^2} \,-\, \frac{a^2 \left(q_1+p+\left(1+\frac{B^2}{a^2}\right)\omega\right)^2}{r^2+a^2} \,-\, \frac{\sin\alpha \cos\alpha}{\sqrt{n(n+2)}}\,\frac{2 B^2 \,\omega \,(q_1+q_2)}{a^2} \\
& \,-\, \frac{\sin^2 \alpha}{n(n+2)}\, \frac{4 B^2\, \omega}{a^4} \left(a^2(p+q_2) +(q_2-q_1) r^2 \right)  \,-\, \frac{\sin^2 \alpha \cos^2 \alpha}{n(n+2)} \,\frac{3 B^4\, \omega^2}{2 a^4} \\
& \,-\, \frac{\sin^4 \alpha}{n^2(n+2)^2} \,\frac{2 B^4\, \omega^2}{a^8} \left(a^4 n (n+2) -2 r^2 (r^2+a^2) \right)\, .
\label{eq:potentialsgenalpha}
\end{split}
\end{equation}
The term proportional to $\cW_n (r,\theta)$ is a complicated function of the coordinates $r$, $\theta$, the mode numbers and $\alpha$. It expresses the failure of separability. However this term vanishes when $\alpha=\frac{\pi}{2}$ and when $\alpha=0$ with $q_1+q_2=0$.  The wave operator is therefore separable in these limits.

\subsubsection{Solution with $\alpha=0$ and $q_1+q_2=0$}

When $\alpha=0$, the contribution from the supercharged superstratum solution is set to zero and the solution is simply the ($2,1,n,q=0$) original superstratum solution \cite{Bena:2017upb}. For this value, $\cW_n(r,\theta)$ takes a simpler form and gives
\begin{equation}
\begin{split}
\cW^{(\alpha=0)}_n(r,\theta) \,=\, & \frac{B^2\,\omega}{2 a^2(r^2+a^2)^2}\Biggl[ \,2 a^4 \left(p-(1+n)\,q_1+(2+n)\,(q_2-\omega)\right) -2 B^2 r^2 \omega\\
&\hspace{2.3cm} -a^2 \left((1+n)B^2 \omega +2 r^2 (q_1-q_2+2 \omega) \right) -2 a^4 (1+n) (q_1 +q_2) \cos2\theta \Biggr] \\
& - \frac{B^4}{4 (r^2+a^2)^2}\, \omega^2\, \left(\frac{r^2}{r^2+a^2} \right)^{1+n}\,.
\label{eq:Wnalpha0}
\end{split}
\end{equation}
The non-separable term is then proportional to $q_1+q_2$ and can be set to zero by considering the modes with $q_2=-q_1$. Then, the scalar wave equation separates and reduces to an angular and radial equations
\begin{equation}
\begin{split}
\frac{1}{r} \partial_r \left( r \left( r^2 + a^2 \right) \partial_r \, K(r)  \right) - V^{(\alpha=0)}_r(r) \,K(r) + \left(\frac{r^2}{r^2+a^2} \right)^n \cW^{(\alpha=0)}_n (r) \, K(r)& \,=\, \lambda\,K(r) \,,
\label{eq:radialeqOrig}\\
\frac{1}{\sin \theta \cos \theta} \partial_\theta \left( \sin \theta \cos \theta \, \partial_\theta \,S(\theta) \right) - \left( \frac{q_1 ^2 }{\sin^2 \theta} + \frac{q_2 ^2 }{\cos^2 \theta} \right) S(\theta) &\,=\, -\lambda S(\theta)\,,
\end{split}
\end{equation}
where we have separated the wave function $H(r,\theta) = K(r)\,S(\theta)$.  The functions $V^{(\alpha=0)}_r(r)$ and  $\cW^{(\alpha=0)}_n (r)$ are given in \eqref{eq:potentialsgenalpha} and \eqref{eq:Wnalpha0} with $\alpha=0$ and $q_1+q_2=0$. The parameter $\lambda$ corresponds to the constant eigenvalue of the Laplacian operator along the $S^3$ and generates  an effective mass in the three-dimensional space-time. Thus, we define the conformal weight of the wave, $\Delta$, via $\lambda \equiv \Delta(\Delta-2)$. Without loss of generality, we consider that $\Delta>1$. The angular equation is solvable and there is only one branch of well-defined solutions \cite{Bena:2017upb,Bena:2018bbd}:
\begin{equation}
S(\theta) \:\propto\:  \left(\sin\theta\right)^{|q_1|}\,  \left(\cos\theta\right)^{|q_2|}\, _2 F_1 \left(-s, 1+ s +|q_1|+|q_2|, |q_2|+1, \cos^{2} \theta \right), 
\label{eq:angularSolution1Orig}
\end{equation}
where $s$ is given by
\begin{equation}
 s=\frac{1}{2}\Big(\Delta -2 -|q_1|-|q_2|\Big).
\label{eq:ldef}
\end{equation}
The solution is regular at $\cos^2\theta = 1$ if and only if
\begin{equation}
~~ \Delta-1 \;\geq\; 1 + |q_1| + |q_2|\,, \qquad\quad q_1,\,q_2 ,\, \Delta, \, s\,\in\, \mathbb{Z} \,, \quad s \geq 0 \,.
\label{eq:angularconstraint}
\end{equation}
Moreover, the regularity of the modes at $r=0$ requires \cite{Tyukov:2017uig}:
\begin{equation}
n_y ~\equiv~  p - \omega  ~\in~ \mathbb{Z}\,.
\label{eq:nydef}
\end{equation}

However, the radial wave equation is not solvable. Here, we will be only interested in the spectrum of normalizable modes and these can be studied using either of the two following approaches:
\begin{itemize}
\item The first method is to apply the WKB approximation by translating the radial equation into a Schr\"odinger problem, as in \cite{GreenSuperstrata}. This is achieved by rescaling $K(r)$  to a new wave function $\Psi(r) \equiv (r^2+a^2)^{-1/2} K(r)$, and change variables $r = a\, e^x$ so that the radial equation is reduced to the standard Schr\"odinger problem:
\begin{equation}
\frac{d^2\Psi}{dx^2}(x) \, -\,  V(x) \Psi(x) \,=\, 0 \,,
\label{eq:shroWE}
\end{equation}
where $V(x)$ is given by
\begin{equation}
V(x) \,\equiv\, \frac{e^{2x}}{1+e^{2x}} \Biggl[\, \lambda +1 +\frac{1}{1+e^{2x}} +V^{(\alpha=0)}_r (a\, e^x) +(1+e^{-2x})^{-n} \,\, \cW^{(\alpha=0)}_n(a\, e^x) \,\Biggr]\,.
\end{equation}
Depending on the range of frequencies and momentum, $\omega$ and $p$, the potential has two ``classical turning points", $x_-$ and $x_+$, satisfying $V(x_\pm)=0$. The spectrum of normalizable modes is then given by the WKB integral depending on the potential between these two points \cite{GreenSuperstrata}
\begin{equation}
\Theta \,\equiv\, \int_{x_-}^{x_+}\, |V(x)|^{\frac{1}{2}}\, dx \,=\, \pi \left( \ell + \frac{1}{2}\right)\,, \quad \ell\in \mathbb{N} \,,
\label{eq:Theta}
\end{equation}
where $\ell$ is an integer label giving the number of the mode.

\item Another approach, developed in \cite{Bena:2018bbd}, allows one to compute, analytically, the spectrum of the $\sqrt{n}$ first normalizable modes by taking $n$ to be large. This limits means that one can neglect those terms in the radial wave equations that are small where the bound-state (normalizable) wave function is large and only become significant in the region in  which the normalizable modes are strongly decaying. More explicitly, the $\ell^{th}$ mode starts to decay at $r\sim \ell \, a$ and all the terms in the radial wave equation that are negligible below this distance have no impact on the profile of the mode. In particular, the terms proportional to $\left(\frac{r^2}{r^2+a^2} \right)^n$ in the radial wave equation are negligible for $r \lesssim \sqrt{n}\,a$. We refer the reader to the Appendix B of \cite{Bena:2018bbd} for a rigorous mathematical proof. Thus, for the $\sqrt{n}$ first normalizable modes, these terms can be neglected. 
\end{itemize}

We first apply the large-$n$ approximation. If we neglect all the terms proportional to $\left(\frac{r^2}{r^2+a^2} \right)^n$ in \eqref{eq:WEgenalpha}, the wave equation separates, even if $q_1+q_2 \neq 0$, into the massive scalar wave equation in a red-shifted AdS$_3$ with constant electric fields and into the angular wave equation of the round $S^3$. According \cite{Bena:2018bbd}, the normalizable modes, at leading order in $\frac{1}{n}$, are in a tower of evenly-spaced frequencies labeled by the mode number $\ell$
\begin{equation}
\omega_\ell \,=\, \eta \, \ell + \delta \,,\quad \ell\in \mathbb{N}  \,.
\label{eq:towermodelargen}
\end{equation}
where $\delta$ is the mass gap, that is, the gap in energy between the ground state and the first excited mode, and $\eta$ is the constant  (to leading order in $\frac{1}{n}$) difference of energies between successive excited modes. In the large-$n$ approximation, one finds \cite{Bena:2018bbd}
\begin{equation}
\begin{split}
\eta &\,=\, \frac{2}{1+\frac{B^2}{2a^2}} \,=\, \frac{4\, j_L}{n_1 n_5}\,,\\
\delta &\,=\, \frac{\Delta +|q_2+n_y| -(q_1+n_y)}{1+\frac{B^2}{2a^2}}\,=\, \frac{2\, j_L}{n_1 n_5} \left( \Delta +|q_2+n_y| -(q_1+n_y) \right)\,.
\end{split}
\label{eq:massgaplargen}
\end{equation}
The minimum mass gap is obtained by considering the bound given by the regularity of the angular wave equation \eqref{eq:angularconstraint}:
\begin{equation}
\delta_{\text{min}} \,=\,  \frac{4\, j_L}{n_1 n_5}  \,.
\end{equation}

When $n$ is not large, we apply the WKB approximation by requiring that $q_2=-q_1$ so that the wave equation is separable. The WKB integral, $\Theta$ \eqref{eq:Theta}, is not integrable analytically and so we perform a numerical analysis and take different values of $\Delta$, $n_y$, $q_1$ and $n$.  We have observed that, for any $n$, the normalizable modes are still in a tower of evenly-spaced frequencies (that depend upon $n$) labeled by the mode number, $\ell$:
\begin{equation}
\omega_\ell \,=\, \eta_n \, \ell + \delta_n \,,\quad \ell\in \mathbb{N}  \,.
\end{equation}
The mass gap, $\delta_n$, and the energy gap between two successive modes, $\eta_n$, differs from their large-$n$ values, \eqref{eq:massgaplargen}, but are of the same order of magnitude. As an illustration, the Fig.\ref{fig:massgap} gives $\eta_n$ and $\delta_n$ as functions of $n$ for a specific value of mode parameters which corresponds to the minimum mass gap value ($\Delta=10$, $q_1=-q_2=4$, $n_y=10$). The graphs corresponding to the ($2,1,n,q=0$) original superstratum are given by the curves in blue. 

To conclude, the  scalar wave perturbations of the ($2,1,n,q=0$) single-mode superstratum solutions have the same properties as the original ($1,0,n,q=0$) single-mode superstratum solutions. The spectra of normalizable modes are linear in the mode number. The minimum mass gap and the energy gap between each mode are of order $4 j_L / (n_1 n_5)$ and match this value for $n$ large. However, one can now investigate the dependence of the spectrum on the internal parameter $\alpha$.  

\subsubsection{Solution with $\alpha=\frac{\pi}{2}$}

When $\alpha=\frac{\pi}{2}$, the contribution from the original superstratum solution is set to zero and the solution is simply the ($2,1,n,q=1$) supercharged superstratum solution \cite{Ceplak:2018pws}. For this value, the scalar wave equation separates and reduces to an angular and radial equations
\begin{equation}
\begin{split}
\frac{1}{r} \partial_r \left( r \left( r^2 + a^2 \right) \partial_r \, K(r)  \right) - V^{(\alpha=\frac{\pi}{2})}_r(r) \,K(r) + \left(\frac{r^2}{r^2+a^2} \right)^n \cW^{(\alpha=\frac{\pi}{2})}_n (r) \, H(r,\theta)& \,=\, \lambda\,K(r) \,,
\label{eq:radialeq}\\
\frac{1}{\sin \theta \cos \theta} \partial_\theta \left( \sin \theta \cos \theta \, \partial_\theta \,S(\theta) \right) - \left( \frac{q_1 ^2 }{\sin^2 \theta} + \frac{q_2 ^2 }{\cos^2 \theta} \right) S(\theta) &\,=\, -\lambda S(\theta)\,,
\end{split}
\end{equation}
where we have separated the wave function $H(r,\theta) = K(r)\,S(\theta)$. The function, $V^{(\alpha=\frac{\pi}{2})}_r(r)$, is given in \eqref{eq:potentialsgenalpha} with $\alpha=\frac{\pi}{2}$, $\cW^{(\alpha=\frac{\pi}{2})}_n (r)$ is only dependent on $r$ but remains a complicated function and $\lambda$ is the constant eigenvalue of the Laplacian operator along the $S^3$. As before, we define the conformal weight of the wave, $\Delta$, via $\lambda \equiv \Delta(\Delta-2)$. The angular wave equation is the same as in the previous section and $S(\theta)$ is  given by \eqref{eq:angularSolution1Orig} with the regularity condition \eqref{eq:angularconstraint}.

The radial wave equation is not solvable analytically. However, the spectrum of normalizable modes can be computed using the large-$n$ method and the WKB approximation as for the ($2,1,n,q=0$) superstratum solutions.

For the large-$n$ limit, we can simply refer to the previous section by noticing that, for the $\sqrt{n}$ first normalizable modes, $$V^{(\alpha=\frac{\pi}{2})}_r(r)\,=\,V^{(\alpha=0)}_r(r) \, \left(1+ \mathcal{O}\left(\frac{1}{n}\right)\right)\, ,\qquad \text{for}\quad  r\lesssim \sqrt{n} a\,.$$ Thus, the wave equation for those modes is essentially the same as in the ($2,1,n,q=0$) superstratum background. At large $n$, the $\sqrt{n}$ first normalizable modes are in a tower of evenly-spaced frequencies labeled by the mode number $\ell$ given in \eqref{eq:towermodelargen} and \eqref{eq:massgaplargen}.

\begin{figure}
\centering
\hspace*{-0.17in}
\includegraphics[width=1.04\textwidth]{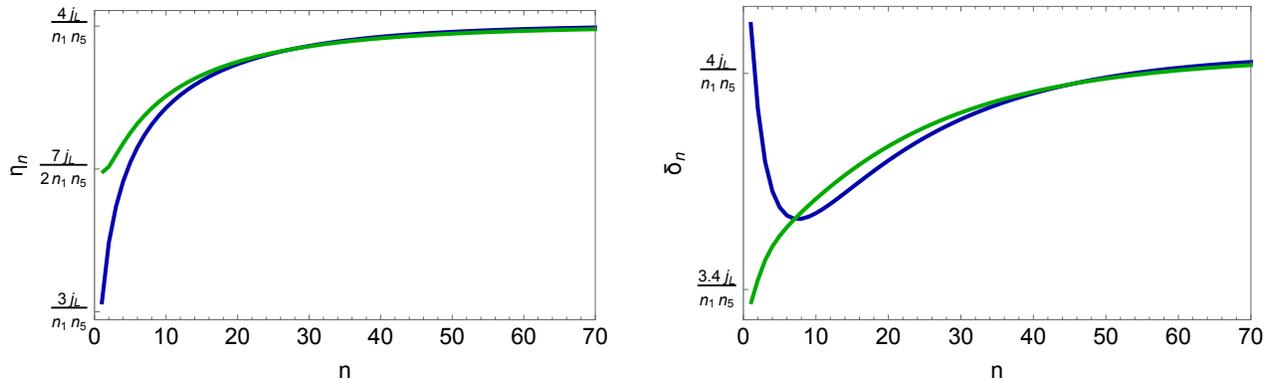}
\caption{The mass gap, $\delta_n$, and the energy difference between two successive normalizable modes, $\eta_n$, of  scalar perturbations in the ($2,1,n,q=0$) original superstratum solution (blue curves) and in the ($2,1,n,q=1$) supercharged superstratum solution (green curves) as functions of $n$. We have chosen particular values of integer mode parameters which corresponds to the minimum mass gap value ($\Delta=10$, $q_1=-q_2=4$, $n_y=10$). However, any other initial values give similar profiles which converge quickly to their large-$n$ values given by \eqref{eq:massgaplargen}.}
\label{fig:massgap}
\end{figure}

For any finite value of $n$, we have performed a similar WKB analysis as for the  ($2,1,n,q=0$) superstratum solutions by numerically integrating the WKB integral, $\Theta$ \eqref{eq:Theta}, for many integer mode parameters $\Delta$, $n_y$, $q_1$ and $q_2$. We have also found that the the spectrum is linear 
\be 
\omega_\ell \,=\, \eta_n \, \ell + \delta_n \,,\quad \ell\in \mathbb{N}  \,.
\ee
where the mass gap, $\delta_n$, and the energy gap between two successive modes, $\eta_n$, are of the same order of magnitude of their large-$n$ values \eqref{eq:massgaplargen}. However, they are significantly different from the mass gap and the energy gap between two successive modes computed at the same $n$ in the ($2,1,n,q=0$) superstratum solutions. As an illustration, the Fig.\ref{fig:massgap} shows the mass gap and the energy gap between two successive modes as functions of $n$ for $\Delta=10$, $q_1=-q_2=4$ and $n_y=10$. The blue plots correspond to the spectrum in ($2,1,n,q=0$) superstratum solutions whereas the green plots correspond to the spectrum in ($2,1,n,q=1$) supercharged superstratum solutions. Their spectra are indeed very close when $n$ is large, they differ significantly for $n$ of order 1.

The mass-gap calculations show that, when $n$ is not large, the spectrum depends on $\alpha$.  It is interesting to consider how one might average states over $\alpha$.  From the CFT perspective the obvious suggestion is to adopt the standard rule in a microcanonical ensemble and average over all  states (having all the same energies and charges) with equal probability.   It would perhaps be more interesting to imagine introducing a small temperature and then the averaging would be biased slightly towards the superstrata with the smallest mass gap.   Either way, averaging over $\alpha$ will create some interesting effects on the spectrum and resonant responses for small values of $n$, but for larger values of $n$, the energy gap will limit to $\frac{4 j_L}{n_1 n_5}$ and the spectrum and response functions will become independent of $\alpha$ and the resonances will become sharp.

\subsubsection{Generic value of $\alpha$}

To have the full picture of the scalar wave perturbation in the ($2,1,n$) hybrid superstrata, one should study the wave equation for generic value of $\alpha$. However, for $\alpha \ne (0,\frac{\pi}{2})$, the wave equation does not separate and must be treated as a two-dimensional second order differential equation. The analysis of normalizable modes in this context requires more sophisticated tools and we will not explore this here. 

On the other hand, one can still apply the large-$n$ approximation for generic value of $\alpha$. Indeed, when $n$ is large, and for the $\sqrt{n}$-first normalizable modes, one can show that
\be 
\begin{split}
V^{(\alpha)}_r(r)&\,=\,V^{(\alpha=0)}_r(r) \, \left(1+ \mathcal{O}\left(\frac{1}{n}\right)\right)\, ,
\qquad \text{for}\quad  r\lesssim \sqrt{n} a\,. \\
V^{(\alpha)}_\theta(\theta)&\,=\,  \left( \frac{q_1 ^2 }{\sin^2 \theta} + \frac{q_2 ^2 }{\cos^2 \theta} \right) \, \left(1+ \mathcal{O}\left(\frac{1}{n}\right)\right)\,.
\end{split}
\ee
Thus, if we neglect the terms proportional to $\left(\frac{r^2}{r^2+a^2} \right)^n$ in \eqref{eq:WEgenalpha}, the wave equation separates to the massive scalar wave equation in a red-shifted AdS$_3$ with constant electric fields and to the angular wave equation of the round $S^3$. The spectrum of $\sqrt{n}$-first normalizable modes at leading order in $\frac{1}{n}$ is the same as the one obtained for $\alpha=0$ and $\alpha=\frac{\pi}{2}$. The modes are in a tower of evenly-spaced frequencies labeled by the mode number $k$ given in \eqref{eq:towermodelargen} and \eqref{eq:massgaplargen}. 

This shows that the $\alpha$-dependence of the spectrum of normalizable modes appears only at the first-order correction in $\frac{1}{n}$ expansion when $n$ is large. The leading order behaviour is essentially determined by the spectrum of a red-shifted AdS$_3$ with constant electric fields. 

When $n$ is of order one,  the wave equation for general $\alpha$ is much richer than the wave equations for $\alpha=0$ or $\frac{\pi}{2}$.  For instance, the angular potential, $V_\theta(\theta)$ \eqref{eq:potentialsgenalpha}, depends  on $\omega$. This is essentially due to the fact that the three-dimensional space determined by $(\theta,\phi,\psi)$ is now a squashed $S^3$. Thus, we  certainly expect that, for generic value of $\alpha$, the spectrum of normalizable modes will no longer have such evenly-spaced modes.

\section{Final comments}
\label{sec:Conclusions}

We have shown how the newly-constructed supercharged superstrata represent an essential missing piece of the puzzle in the construction of supergravity duals of  generic superstratum states (\ref{multistates}).  Such states, in principal,  allow arbitrary occupation numbers, $N_{k_i,m_i,n_i,q_i}$, (subject to (\ref{sbudget})) and hence the  fundamental Fourier coefficients  in holographic dual should be freely choosable (subject to the corresponding supergravity constraints that generalize (\ref{SGbudget1})).   This was a problem for the original superstratum because the interaction of two generic modes, $(k_1,m_1,n_1)$ and $(k_2,m_2,n_2)$, led to dangerous, uncanceled high-frequency sources that were proportional to $b_4 b_5$  (see  (\ref{coiff1})).  While it has not been rigorously proved that these sources must necessarily lead to singular supergravity solutions, these sources were, at the least, a significant unsolved problem for the original superstrata.   Indeed, they appear to cover a  restricted range of CFT states and they suggest that such supertstrata may only be functions of two variables.  

As we saw in Section \ref{ss:modes}, and particularly from  (\ref{coiff1}), the introduction of supercharged modes solves the problem entirely: the unwanted terms can be trivially coiffured away as part of the coiffuring of the supercharged modes.  This means that superstrata fluctuations involve generic functions of three variables and can describe the full range of states in (\ref{multistates}). 

One of the structural features of the solution generating methods that led to the standard coiffuring relations for superstrata based on $|00\rangle$ strands is that they involve using only {\it finite} Fourier series for all fluctuating terms.   It is important to remember that there are other ways to resolve the coiffuring constraints, and this could involve  a much broader range of solutions using {\it infinite} Fourier series. 

There are several examples of more general classes of coiffuring. First, the smoothness of supertubes with varying charge densities (see, for example, \cite{Bena:2010gg}) and coiffuring relations for black rings with non-trivial charge densities \cite{Bena:2014rea}  involve constraints on the charge densities at the supertube or at the horizon.  For such configurations in the supergravity theory used in this paper, there are three electric charge densities, $\rho_1(v)$, $\rho_2(v)$ and $\rho_4(v)$, and an angular momentum charge density,  $\rho_J(v)$,  where $v$ is the null coordinate along the tube or horizon.   Smoothness typically places two quadratic constraints on these densities.  Indeed, one such constraint  involves requiring that the quadratic $\rho_1(v) \rho_2(v) + \rho_4(v)^2$ be independent of $v$.    If  $\rho_1(v)$ does not vanish, one can render $\rho_1(v) \rho_2(v) + \rho_4(v)^2$  independent of $v$ by taking $\rho_4 \equiv 0$ and $\rho_2(v) = (\rho_1(v))^{-1}$, as was done in \cite{Bena:2010gg}.    For superstrata, one can include excitations of $(Z_2, \Theta_1)$, as in  \cite{Bena:2016agb}, and this also allows more general styles of coiffuring.   When the dangerous high-frequency terms in the original superstrata became evident, it was suspected that a more complicated set of coiffuring relations, perhaps involving infinite Fourier series might be needed.   
 
The disadvantage of such a ``pure supergravity'' resolution of singularities is that one loses contact with the established holographic dictionary.  Indeed, for the supergraviton gas states based on  $|00\rangle$ strands, we know precisely how the superstratum states appear, at leading order, as finite Fourier series in $\Theta_4$.  It is therefore extremely nice to be able to coiffure the high-frequency  sources in a manner that is faithful to the known holographic dictionary.   The result is especially gratifying in that it does not lead to constraints on the leading Fourier coefficients in $\Theta_4$ and, instead, leads to testable predictions about the three-point functions in the dual CFT states.    It would also be very nice to develop holographic dictionaries for other, more complicated coiffuring techniques. Indeed, the results of  \cite{Bena:2016agb} suggest that such techniques will lead to the holographic duals of supergraviton states based on other types of strand.

There is another interesting issue in the dual CFT that we glossed over in the introduction:  what is the holographic dual of our hybrid superstrata?  In writing  (\ref{multistates}), we simply worked by analogy with the dictionary developed in earlier papers on superstrata, like \cite{Bena:2015bea}.  Because the original (bosonic) superstratum modes and the supercharged superstratum with the same mode numbers are so similar, and so nearly degenerate, one might reasonably wonder if there might be some other possible identification of these holographically dual states.    However, it is important to remember that the holographic identifications, like (\ref{multistates}), are really place-holders, representing the peak of a coherent state.  (For a discussion of this, see, for example \cite{Giusto:2015dfa}.) Thus even non-hybridized superstrata involve smearing of classical states into a coherent wave-functions.  Incorporating the hybridized superstrata into such a picture simply means that the coherent state will spread across more CFT states, but once again one will have to identify the peak of this coherent spread and indeed one expects\footnote{We are grateful to Rodolfo Russo for his explanations of this issue.} that this will occur at (\ref{multistates}) with the occupation numbers related to the Fourier coefficients via (\ref{NandFCnorms}).

Another broad thread in this work is how the resolution of singularities in supergravity is dual to the correct description of the infra-red physics in the CFT.  We remarked on this in the introduction when we discussed how holographic RG flows like that of confining QCD.  Another way to think about this is that, in black-hole physics and in holographic renormalization group flows, singularities arise when one does not include the essential degrees of freedom needed to resolve singularities. The microstate geometry progamme {\it exists} because string theory and supergravity have enough degrees of freedom to resolve black-hole singularities.  Moreover, as the microstate geometry program evolves we see this principle re-iterated time and again.   Early attempts to make the original superstrata involved using finite Fourier series in only the pairs $(Z_2, \Theta_1)$ and  $(Z_1, \Theta_2)$ (with $(Z_4 =0 , \Theta_4 =0)$).  Such superstrata could not be constructed because of the uncanceled singularities coming from high-frequency sources.  However, string scattering calculations \cite{Giusto:2011fy,Giusto:2012gt, Giusto:2012jx, Shigemori:2013lta, Giusto:2013bda} showed that  a generic amplitude that involved sourcing $Z_I$ and  $\Theta_I$ (for $I=1,2$) would necessarily activate the degrees of freedom that source  $(Z_4, \Theta_4)$.  It was thus inconsistent with string scattering to suppress this degree of freedom.  Indeed, once $(Z_4, \Theta_4)$ was included, coiffuring became simple and the result was the breakthrough in \cite{Bena:2015bea}.  In this paper we see it again: we have shown that the original superstrata have singular interactions that can be obviated allowing the excitation supercharged superstratum modes.

The presence of free parameters, unconstrained by bulk thermodynamic charges, is perhaps the most interesting feature of our hybrid single-mode superstrata.  One can think of such parameters as mapping out tiny corners of the phase space of the dual black hole.  Thus one can consider trying to average probes, like the results of \cite{GreenSuperstrata}, over such families of states and perhaps trying to average-out some of the sharp, classical features of the geometry.  The question then becomes how one should perform this averaging, and what is the measure?  For BPS states, in which the temperature vanishes, this returns us to the issues raised above:  what is the mixing and how are the coherent states peaked.  Once one understands that, one can look at the averaging process in terms of averaging over CFT states.

If one breaks the supersymmetry and introduces a small temperature then the results of Section \ref{sec:example} suggest that degeneracy of the supergravity backgrounds may be broken: the mass-gaps of the supercharged and original superstrata are slightly different and so Boltzmann weights will give a slight bias to a particular class of geometries.  

While the single-mode hybrid geometries constructed in this paper are relatively simple, they are, perhaps, not as interesting as generic multi-mode superstrata.  This is precisely because the single-mode hybrids are so degenerate.  It was shown in  \cite{Bena:2016ypk,Bena:2017xbt} that the momentum-wave structure that causes the transition from BTZ-like geometry to the cap of the microstate geometry is approximately localized at $r \sim \sqrt{\frac{n}{k}} \, a$ (at least for $m=0$), and so different mode numbers introduce different scales into the geometry. Probes and Green functions will reflect and echo from these structures, and even small differences of scale can create interference that can wash-out coherent signals from individual geometries.  Thus, we expect the idea of averaging over multi-mode superstrata to produce a much richer story.  It is therefore extremely encouraging to know that there are no longer any obvious obstructions to constructing the supergravity duals of the generic superstratum state.

\vspace{0.8cm}

\section*{Acknowledgments}
\vspace{-2mm}
We would like to thank  Rodolfo Russo and Masaki Shiegomori for valuable discussions.  The work of NW was supported in part by ERC Grant number: 787320 - QBH Structure and by DOE grant DE-SC0011687. The work of PH was supported by an ENS Lyon grant and by the ANR grant Black-dS-String  ANR-16-CE31-0004-01.

\newpage

\begin{adjustwidth}{-1mm}{-1mm} 

\bibliographystyle{utphys}      

\bibliography{microstates}       

\end{adjustwidth}

\end{document}